\documentclass[review,1p,authoryear]{elsarticle}

\usepackage{fullpage}
\usepackage{amsmath,amssymb,amsthm}
\usepackage{stmaryrd}
\usepackage{float}
\usepackage{subfigure}
\usepackage{graphicx}
\usepackage{color}
\usepackage{tabu}
\usepackage[algoruled,vlined,boxed,longend,english,lined,boxed,commentsnumbered]{algorithm2e} % para inserir pseudocodigos	

\newtheorem{theorem}{Theorem}

\newtheorem{remark}[theorem]{Remark}

\newcommand{\sip} {\!\cdot\!}
\newcommand{\dip} {\!:\!}
\newcommand{\GM}{\Gamma_{\mbox{\tiny{\!$m$}}}}
\newcommand{\GN}{\Gamma_{\mbox{\tiny{\!N}}}}
\newcommand{\GD}{\Gamma_{\mbox{\tiny{\!D}}}}
\newcommand{\hh}{\hspace*{0.7pt}}
\newcommand{\si}{_{{\mbox{\tiny{$(i)$}}}}}
\newcommand{\sj}{_{{\mbox{\tiny{$(j)$}}}}}
\newcommand{\ssi}{_{{\mbox{\tiny{$(1)$}}}}}
\newcommand{\sxi}{_{\!{\mbox{\tiny{$\boldsymbol{\xi}$}}}}}

\begin{document}
 
\begin{frontmatter}

\title{On the full-waveform inversion of seismic moment tensors}

\author[SUBC] {Alan A.S. Amad}
\ead{a.a.s.amad@swansea.ac.uk}

\author[LNCC]{Antonio A. Novotny}
\ead{novotny@lncc.br}

\author[UMN]{Bojan B. Guzina\corref{cor1}}
\ead{guzin001@umn.edu}

\address[SUBC]{Zienkiewicz Centre for Computational Engineering, College of Engineering, Swansea University Bay Campus, Swansea, Wales, UK}

\address[LNCC]{Laborat\'{o}rio Nacional de Computa\c{c}\~{a}o Cient\'{\i}fica LNCC/MCT, Coordena\c{c}\~{a}o de M\'{e}todos Matem\'{a}ticos e Computacionais, Av. Get\'{u}lio Vargas 333, 25651-075 Petr\'{o}polis - RJ, Brasil}

\address[UMN]{Department of Civil, Environmental, and Geo- Engineering, University of Minnesota, Minneapolis, MN 55455, US}

\cortext[cor1]{Corresponding author.}

\date{\today}
%%%%%%%%%%%%%%%%%%%%%%%%%%%%%%%%%%%%%%%%%%%%%%%%%%%%%%%%%%%%%%%%%%%%%%%%%%%%%%%%%%%%%%%%%%%%%%%%%%%%%%%%%%%%%%%%%%%%%%%%%%
\begin{abstract}
In this work, we propose a full-waveform technique for the spatial reconstruction and characterization of (micro-) seismic events via joint source location and moment tensor inversion. The approach is formulated in the frequency domain, and it allows for the simultaneous inversion of multiple point-like events. In the core of the proposed methodology is a grid search for the source locations that encapsulates the optimality condition on the respective moment tensors. The developments cater for compactly supported elastic bodies in $\mathbb{R}^2$; however our framework is directly extendable to inverse (seismic) source problems in $\mathbb{R}^3$ involving both bounded and unbounded elastic domains. A set of numerical results, targeting laboratory applications, is included to illustrate the performance of the inverse solution in situations involving: (i)~reconstruction of multiple events, (ii)~sparse (pointwise) boundary measurements, (iii)~``off-grid" location of the micro-seismic events, and (iv) inexact knowledge of the medium's elastic properties.
\end{abstract}


\end{frontmatter}

% --------------------------------------------------------------------------
\section{Introduction} 
% --------------------------------------------------------------------------
\noindent Seismic and micro-seismic source characterization is a keen area of research in geophysics, engineering, hydrocarbon production, and materials science due to its central role in the understanding of earthquake and faulting processes~\citep{Shea2009}; monitoring of mines, highway bridges, and offshore platforms~\citep{Koer1981}; tracking the progress of hydraulic fracturing~\citep{Baig2010}, and investigating the failure of brittle materials~\citep{Gros2008}. Generally speaking any (micro-) seismic source, interpreted as a sudden material failure, can be characterized by its spatial support, temporal variation, and the underpinning failure mechanism. In situations when the extent of a material failure is small relative to the remaining length scales in the problem (e.g. seismic wavelengths and source-receiver distances), the seismic source can be interpreted as a point source~\citep{Scru1985,Jost1989}; a hypothesis that is implicitly assumed hereon. In this setting, the accepted continuum mechanics description of a seismic source is given by a linear combination of force dipoles~\citep{Aki2002} whose weights are specified in terms of the so-called \emph{seismic moment tensor}~\citep{Gilb1971}; a second-order tensorial quantity whose accurate reconstruction from remote wavefield measurements is the lynchpin of seismic source characterization. 

Transcending the classical approaches to moment tensor inversion in laboratory~\citep{Scru1985} and geophysical~\citep{Jost1989} environments that rely on prior knowledge of the source location and possibly other simplifying assumptions (e.g. far field hypothesis), recent attempts at seismic source characterization are increasingly based on the full waveform analysis of multi-axial seismic observations~\citep{Cesc2015}. In general, the latter can be pursued either via time- or frequency-domain approaches. As an example of the former class of inverse solutions, \cite{Song2011} deploy grid search for the source location -- aiming to minimize the $L^2$ misfit between the observed and synthetic waveforms, followed by a least-squares solution for the moment tensor that relies on an \emph{a priori} premise of the source time function. In~\cite{Sjor2014}, on the other hand, the investigators pursue simultaneous inversion for the source location, moment tensor, and two-parameter source time function via nonlinear minimization of the germane $L^2$ waveform misfit, aided by adjoint-field sensitivity estimates. In recent years, studies~\citep{Baza2015,Kawa2008} have demonstrated the utility of time reversal methods as a viable (time- or frequency-domain) alternative for exposing the seismic source location. With the latter information at hand, a full-waveform reconstruction of the moment tensor, including the underpinning source time function, can be conveniently pursued in the frequency domain~\citep{Cesc2008} by solving the underpinning linear system of equations. 

A common thread to the above and related inverse source analyses entails (i) the fundamental premise of a \emph{synchronous seismic source}, where all components of its moment tensor share the same time dependence (given by the source time function); and (ii) the assumption of \emph{a single} seismic (point) source, precluding the possibility that two events -- originating from distinct locations -- may overlap in time. To provide an alternative to the foregoing analyses that is free of such impediments, this work deals with spatial reconstruction and characterization of micro-seismic events in the frequency domain from pointwise wavefield measurements, where both real and imaginary parts of the associated moment tensors are fully reconstructed. Since the inverse problem at hand is (as expected) ill-posed, the idea is to rewrite it as an optimization problem in which a functional measuring the misfit between synthetic and observed waveforms is minimized with respect to a set of admissible point sources representing the hidden faults. The necessary optimality conditions are derived in the spirit of the topological derivative method \citep{NovotnyBook2013,NovotnyBook2019} which, in this context, consists in exposing the perturbation of the functional as a quadratic function of the germane moment tensor components. Then, the resulting expansion is trivially minimized with respect to the sought source parameters, leading to a non-iterative reconstruction algorithm that is initial guess-free and robust with respect to perturbations of sensory data. We test the proposed technique via numerical experiments designed to examine its performance under a variety of source, sensing, and uncertainty scenarios.

The paper is organized as follows. The germane (frequency-domain) forward problem and affiliated inverse problem, targeting the locations and moment-tensor ``strengths'' of micro-seismic events from the observed acoustic emission data, are described in Section~\ref{sec:model}. In Section~\ref{sec:expansion} the germane cost functional, measuring the~$L^2$ misfit between the synthetic and sensory data, is expanded with respect to the set of admissible source densities. The resulting expansion is used to devise a novel reconstruction algorithm presented in Section~\ref{sec:algorithm}. A set of numerical experiments examining the effectiveness of the proposed reconstruction algorithm is provided in Section~\ref{sec:umerics}.

%Finally, the paper ends in Section~\ref{sec:conclu} with some concluding remarks and further discussion.

%If the source function is not known or the hypothesis of source synchronicity is forgone, the frequency domain analysis is adopted~\citep{Gilb1973}. 

% --------------------------------------------------------------------------
\section{Inverse problem} \label{sec:model}
% --------------------------------------------------------------------------

\begin{figure}[htbp]
\begin{center}
\includegraphics[width=0.48\linewidth]{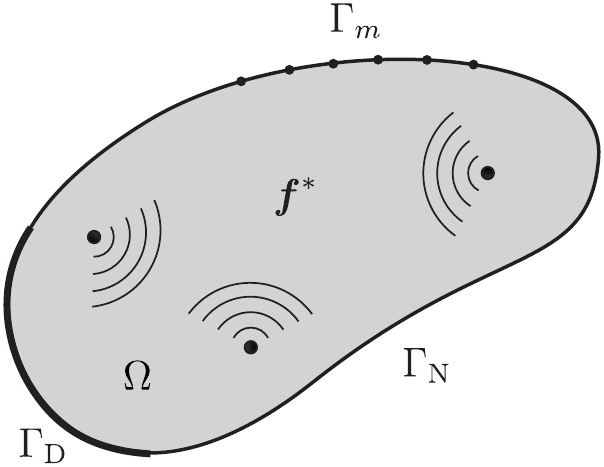} 
\end{center} \vspace*{-5mm}
\caption{Problem setting.}
\label{fig:dpole}
\end{figure}

\noindent Consider a bounded elastic body $\Omega \subset\mathbb{R}^{2}$ endowed with Lipschitz boundary $\partial\Omega$, mass density~$\rho$, and fourth-order elasticity tensor~$\boldsymbol{C}$. For further reference, let~$\GN\subset\Gamma$ and~$\GD=\partial\Omega\setminus\GN$ denote respectively the parts of~$\partial\Omega$ subjected to homogeneous Neumann and Dirichlet boundary conditions. In this setting, we are interested in the \emph{inverse source problem} of reconstructing the source density~$\boldsymbol{f}^*$ such that
\begin{equation}
\left\{
\begin{array}{rllcl}
-\nabla\sip(\boldsymbol{C} \dip \nabla\boldsymbol{u}) - \rho \hh \omega^2 \boldsymbol{u} & = & \boldsymbol{f}^{\ast} & \text{in} & \Omega , \\
\boldsymbol{u} & = & \boldsymbol{u}^{\ast} & \text{on} & \GM ,\\
\boldsymbol{u} & = & \boldsymbol{0} & \mbox{on} & \GD , \\
\boldsymbol{n} \sip (\boldsymbol{C} \dip \nabla\boldsymbol{u}) & = & \boldsymbol{0} & \text{on} & \GN,
\end{array}
\right.
\label{eq:strongZ}
\end{equation}
where $\boldsymbol{u}: \Omega \rightarrow \mathbb{C}^{2}$ is the elastodynamic displacement field; $\omega$ denotes the frequency of wave motion; $\boldsymbol{n}$ is the unit outward normal on~$\partial{\Omega}$; $\GM\!\subset\GN$ is the measurement surface; and~$\boldsymbol{u}^*$ are the \emph{``acoustic emission'' data} from which we aim to resolve~$\boldsymbol{f}^*$, see Fig.~\ref{fig:dpole}. Hereon, we assume the elastic body~$\Omega$ to be homogeneous and isotropic, in which case the elasticity tensor reads  
\begin{equation}
\boldsymbol{C} = 2 \mu\hh \boldsymbol{I}_4 + \lambda \boldsymbol{I}_2 \otimes \boldsymbol{I}_2,
\label{eq:TensorC}
\end{equation}
where~$\lambda$ and~$\mu$ are the Lam\'{e} moduli, and~$\boldsymbol{I}_n$ is the symmetric $n$th-order identity tensor. 

In the spirit of acoustic emission problems, we next describe the source density $\boldsymbol{f}^{\ast}$ via superposition of a finite number of dipoles; specifically, we assume that $\boldsymbol{f}^{\ast}\in C_\delta(\Omega)$, where 
\begin{equation} 
C_\delta(\Omega) = \Big\{ \boldsymbol{f} : \Omega \rightarrow \mathbb{C}^{2}\;|\;
  \boldsymbol{f}(\boldsymbol{x}) = \sum_{i = 1}^{N}\boldsymbol{M}\!\si \sip \nabla \sxi \hh \delta(\boldsymbol{x} - \boldsymbol{\xi})_{|_{\boldsymbol{\xi}=\boldsymbol{\xi}\si}}\Big\}.
\label{eq:Cdelta}
\end{equation}
Here, $\delta(\boldsymbol{\cdot})$ is the two-dimensional Dirac delta function; $N$ denotes the number of point sources located at $\boldsymbol{\xi}\si\in\Omega$ ($i=\overline{1,N}$), and \mbox{$\boldsymbol{M}\!\si\in \mathbb{C}^{2\times 2}$} is a (symmetric) \emph{seismic moment tensor} characterizing the~$i$th point source. For completeness, we recall the continuum mechanics definition~\citep{Aki2002} of the seismic moment tensor as
\begin{equation}
\boldsymbol{M} \:=\: a \,\llbracket\boldsymbol{u}\rrbracket \otimes \boldsymbol{\eta} : \boldsymbol{C},
\label{mtensor}
\end{equation}
where~$a$ is the area of a newly created micro-fracture (giving rise to the acoustic emission) whose unit normal is denoted by~$\boldsymbol{\eta}$, and $\llbracket\boldsymbol{u}\rrbracket$ is the average displacement jump across the micro-fracture. On the basis of~\eqref{eq:Cdelta}, we write the sought source density satisfying~\eqref{eq:strongZ} as
\begin{equation} 
\boldsymbol{f}^\ast(\boldsymbol{x}) = \sum_{i = 1}^{N^\ast} \boldsymbol{M}\si^\ast \sip \nabla\sxi \hh \delta(\boldsymbol{x} - \boldsymbol{\xi})_{|_{\boldsymbol{\xi}=\boldsymbol{\xi}\si^*}} 
\label{truef}
\end{equation}

\begin{remark}
To establish a clear connection of the above time-harmonic setup with physical applications, we denote by~$\boldsymbol{\mathcal{M}}^*\si(t)$ the temporal record of a moment tensor describing the \mbox{(micro-)} seismic event occurring at~$\boldsymbol{\xi}\si^*$, and we assume (without loss of generality) that $t=0$ marks the onset of the event. In this case, we have
\begin{equation} \label{zerotime}
\begin{array}{ll} \boldsymbol{\mathcal{M}}^*\si(t) = \boldsymbol{0}, & t<0, \\*[1mm]
\boldsymbol{\mathcal{M}}^*\si(t) \neq \boldsymbol{0}, & t\to\infty
\end{array}
\end{equation}
due to creation of a permanent dislocation at~$\boldsymbol{\xi}\si^*$. As a result, the components of $\boldsymbol{\mathcal{M}}^*\si(t)$ are not amenable to the Fourier transform. However their temporal derivatives are, in which case the moment tensors in~\eqref{truef} should be interpreted as 
\[
\boldsymbol{M}\!\si^{\; *} \,=\, \boldsymbol{M}\!\si^{\; *}(\omega) \,=\, \frac{1}{\mathfrak{i}\hh\omega} \hh \mathcal{F}\Big[\frac{\text{d}}{\text{d}t}\boldsymbol{\mathcal{M}}^*\si(t)\Big](\omega),
\]
where~$\mathcal{F}[\cdot]$ denotes the Fourier transform and $\mathfrak{i}=\sqrt{-1}$, see~\cite{Rice1980} for an in-depth discussion. 
\end{remark}

\begin{remark}
Most of the existing approaches to moment tensor inversion are based on the assumption of a synchronous seismic source, which states that all components of the moment tensor $\boldsymbol{\mathcal{M}}^*\si(t)$ carry the same time dependence -- referred to as the source time function \citep{Song2011}. In this work, we implicitly dispense with such hypothesis; as examined in~\cite{Jost1989}, this is one of the key advantages afforded by the frequency-domain inversion of seismic moment tensors.  
\end{remark}

Let us rewrite the inverse problem \eqref{eq:strongZ} as an optimization problem. The associated $L^2$ functional to be minimized in $C_\delta(\Omega)$ is given by 
\begin{equation}
\mathcal{J}(\boldsymbol{u}) :=  \dfrac{1}{2} \int_{\GM} \left(\boldsymbol{u} - \boldsymbol{u}^\ast \right) \cdot \overline{\left(\boldsymbol{u} -
\boldsymbol{u}^\ast\right)},
\label{eq:shfuncOri}
\end{equation}
where $\boldsymbol{u} : \Omega \rightarrow \mathbb{C}^{2}$ solves the boundary value problem 
\begin{equation}
\left\{
\begin{array}{rllcl}
-\nabla\sip(\boldsymbol{C} \dip \nabla\boldsymbol{u}) - \rho \hh \omega^2 \boldsymbol{u} & = & \boldsymbol{f} & \text{in} & \Omega , \\
\boldsymbol{u} & = & \boldsymbol{0} & \mbox{on} & \GD , \\
\boldsymbol{n} \sip (\boldsymbol{C} \dip \nabla\boldsymbol{u}) & = & \boldsymbol{0} & \text{on} & \GN,
\end{array}
\right.
\label{eq:strongOri}
\end{equation}
for a \emph{trial} source term $\boldsymbol{f}\in C_\delta(\Omega)$. In this setting, the relevant optimization problem can be stated as 
\begin{equation}
\underset{\boldsymbol{f} \in C_\delta(\Omega)}{\text{Minimize}} \; \mathcal{J}(\boldsymbol{u}) ~ \text{subject to \eqref{eq:strongOri}}.
\label{eq:OptProb}
\end{equation}

% --------------------------------------------------------------------------
\section{Sensitivity Analysis} \label{sec:expansion}
% --------------------------------------------------------------------------

\noindent The next step is to minimize the misfit functional \eqref{eq:shfuncOri} with respect to the set of admissible solutions \eqref{eq:Cdelta}. In order to evaluate the germane sensitivities of this functional, the idea is to perturb the trial source term $\boldsymbol{f} \in C_\delta(\Omega)$ in \eqref{eq:strongOri} by a fixed number, $N$, of point sources with arbitrary locations and \textcolor{black}{generic} moment tensors as
\begin{equation}
\boldsymbol{f}_{\!p}(\boldsymbol{x}) = \boldsymbol{f}(\boldsymbol{x}) + \sum_{i=1}^{N}\boldsymbol{M}\si \sip \nabla \! \si \delta(\boldsymbol{x}),
\label{eq:fdelta}
\end{equation}
where $\nabla \! \si \delta(\boldsymbol{x}) := \nabla \sxi \delta(\boldsymbol{x} - \boldsymbol{\xi})_{|_{\boldsymbol{\xi} = \boldsymbol{\xi}\si}}$, \textcolor{black}{ and \mbox{$\boldsymbol{M}\!\si\!\in \mathbb{C}^{2\times 2}$} are symmetric}. \textcolor{black}{Hereon, we refer to $\boldsymbol{f}_{\!p}\in C_\delta(\Omega)$ as a perturbed source, and we seek to reconstruct~$\boldsymbol{M}\!\si$ (for a given trial set~$\boldsymbol{\xi}\si$, $i=\overline{1,N}$) by direct inversion}. On the basis of \eqref{eq:strongOri} and \eqref{eq:fdelta}, we can introduce the forward solution $\boldsymbol{u}_{p}$ as that solving 
\begin{equation}
\left\{
\begin{array}{rllcl}
-\nabla\sip(\boldsymbol{C} \dip \nabla\boldsymbol{u}_{p}) - \rho \hh \omega^2 \boldsymbol{u}_{p} & = & \boldsymbol{f}_{\!p} & \text{in} & \Omega, \\
\boldsymbol{u}_{p} & = & \boldsymbol{0} & \mbox{on} & \GD, \\
\boldsymbol{n} \sip (\boldsymbol{C} \dip \nabla\boldsymbol{u}_{p}) & = & \boldsymbol{0} & \text{on} & \GN,
\end{array}
\right.
\label{eq:strongPer}
\end{equation}
which gives rise to the perturbed cost functional 
\begin{equation}
\mathcal{J}(\boldsymbol{u}_{p}) = \dfrac{1}{2}\int_{\GM} \left( \boldsymbol{u}_{p} - \boldsymbol{u}^\ast\right) \cdot \overline{\left(
\boldsymbol{u}_{p} - \boldsymbol{u}^\ast \right)}.
\label{eq:shfuncPer}
\end{equation}

Assuming a sufficient number of ``micro-seismic'' source locations $\boldsymbol{\xi}\si$ ($i=\overline{1,N}$), we are interested in obtaining the variation of \eqref{eq:shfuncOri} with respect to the components of the moment tensor $\boldsymbol{M}\si$ at each location. To facilitate the analysis, one may decompose $\boldsymbol{M}\si$ into the real and imaginary parts as
\begin{equation}
\boldsymbol{M}\si = \boldsymbol{A}\si + \mathfrak{i}\hh \boldsymbol{B}\si, \qquad \boldsymbol{A}\si, \boldsymbol{B}\si \in \mathbb{R}^{2\times 2}.
\label{eq:AB}
\end{equation}
Using Einstein summation notation over repeated indexes $k,l=\overline{1,2}$, we can further write 
\begin{equation}
\boldsymbol{A}\si \nabla \! \si \delta(\boldsymbol{x}) = A^{kl}\si (\boldsymbol{e}_k \otimes \boldsymbol{e}_l) \nabla \! \si \delta(\boldsymbol{x}) \quad \text{and} \quad
\boldsymbol{B}\si \nabla \! \si \delta(\boldsymbol{x}) = B^{kl}\si (\boldsymbol{e}_k \otimes \boldsymbol{e}_l) \nabla \! \si \delta(\boldsymbol{x}) \;,
\end{equation}
where $\boldsymbol{e}_k$ and $\boldsymbol{e}_l$ are the unit vectors of the reference Cartesian frame, and $A\si^{kl}$ (resp.~$B\si^{kl}$) are the  components of $\boldsymbol{A}\si$ (resp.~$\boldsymbol{B}\si$). With such definitions, the solution of \eqref{eq:strongPer} can be conveniently decomposed as 
\begin{eqnarray}
\boldsymbol{u}_{p}(\boldsymbol{x}) \:=\: \boldsymbol{u}(\boldsymbol{x}) +
\sum_{i=1}^{N} \big(A^{kl}\si \,\boldsymbol{p}^{kl}\si(\boldsymbol{x}) + B^{kl}\si\, \mathfrak{i}\hh \boldsymbol{p}^{kl}\si(\boldsymbol{x})\big) 
\label{eq:Udelta}
\end{eqnarray}
where $\boldsymbol{p}^{kl}\si$ solve the canonical boundary value problems
\begin{equation}
\left\{
\begin{array}{rllcl}
-\nabla\sip(\boldsymbol{C} \dip \nabla \boldsymbol{p}^{kl}\si) - \rho \omega^2 \boldsymbol{p}^{kl}\si & = &
(\boldsymbol{e}_k \otimes \boldsymbol{e}_l) \nabla \! \si \delta & \text{in} & \Omega\;, \\
\boldsymbol{p}^{kl}\si & = & \boldsymbol{0} & \text{on} & \GD \;, \\
\boldsymbol{n} \sip (\boldsymbol{C} \dip \nabla \boldsymbol{p}^{kl}\si) & = & \boldsymbol{0} & \text{on} & \GN, 
\end{array}
\right.
\label{eq:strongRkl}
\end{equation}
for $k,l=\overline{1,2}$. Here it is useful to note that, thanks to ansatz \eqref{eq:Udelta}, canonical problems \eqref{eq:strongRkl} are independent of the components $A^{kl}\si$ and $\mathfrak{i} B^{kl}\si$ of the moment tensor $\boldsymbol{M}\si$ in \eqref{eq:AB}. Now we have all elements needed to evaluate the variation of functional~\eqref{eq:shfuncOri} with respect to $A^{kl}\si$ and $\mathfrak{i} B^{kl}\si$. Specifically, on substituting \eqref{eq:Udelta} in \eqref{eq:shfuncPer}, we obtain
\begin{multline}
\mathcal{J}(\boldsymbol{u}_{p}) =
\mathcal{J}(\boldsymbol{u}) +
\int_{\GM} \sum_{i=1}^{N} A^{kl}\si \,\Re \left\{\boldsymbol{p}^{kl}\si \cdot \overline{(\boldsymbol{u} - \boldsymbol{u}^\ast)} \right\} +
\int_{\GM} \sum_{i=1}^{N} B^{kl}\si \,\Im \left\{-\boldsymbol{p}^{kl}\si \cdot \overline{(\boldsymbol{u} - \boldsymbol{u}^\ast)} \right\} \\ +
\dfrac{1}{2} \int_{\GM} \sum_{i=1}^{N} \sum_{j=1}^{N} A^{kl}\si A^{mn}\sj \, \boldsymbol{p}^{kl}\si \cdot \overline{\boldsymbol{p}^{mn}\sj} +
\dfrac{1}{2} \int_{\GM} \sum_{i=1}^{N} \sum_{j=1}^{N} B^{kl}\si B^{mn}\sj \, \boldsymbol{p}^{kl}\si \cdot \overline{\boldsymbol{p}^{mn}\sj}, 
\label{eq:expansion}
\end{multline}
assuming implicit summation over repeated indexes $k,l,m,n=\overline{1,2}$.

For a systematic treatment of~\eqref{eq:expansion}, we next introduce the vector of trial source locations 
\begin{equation}
\boldsymbol{z} = \big(\boldsymbol{\xi}_{(1)},\boldsymbol{\xi}_{(2)},\ldots,\boldsymbol{\xi}_{(N)}\big) \in \mathbb{R}^{2N}
\label{bzdef}
\end{equation}
and the affiliated ``strength" vectors 
\begin{eqnarray}
&&\boldsymbol{a} = \big(\boldsymbol{\alpha}_{(1)},\boldsymbol{\alpha}_{(2)}, \cdots, \boldsymbol{\alpha}_{(N)}\big)\in \mathbb{R}^{3N} \\
&&\boldsymbol{b} = \big(\boldsymbol{\beta}_{(1)},\boldsymbol{\beta}_{(2)}, \cdots, \boldsymbol{\beta}_{(N)}\big)\in \mathbb{R}^{3N}
\end{eqnarray}
collecting the respective components of $\boldsymbol{M}\si$, where 
\begin{eqnarray} \notag
&&\boldsymbol{\alpha}_{(i)} = (A^{11}\si, \hh A^{22}\si, \hh A^{12}\si = A^{21}\si), \\
&&\boldsymbol{\beta}_{(i)} = (B^{11}\si, \hh B^{22}\si, \hh B^{12}\si = B^{21}\si). \notag
\end{eqnarray}
With such definitions, the residual in \eqref{eq:expansion} can be rewritten more compactly as 
\begin{eqnarray}
\Psi(N,\boldsymbol{z},\boldsymbol{a},\boldsymbol{b}) &:=& \mathcal{J}(\boldsymbol{u}_{p}) - \mathcal{J}(\boldsymbol{u}) \\
&=&
\boldsymbol{g} \cdot \boldsymbol{a} + \dfrac{1}{2} \boldsymbol{G}\, \boldsymbol{a} \cdot \boldsymbol{a} +
\boldsymbol{h} \cdot \boldsymbol{b} + \dfrac{1}{2} \boldsymbol{G}\, \boldsymbol{b} \cdot \boldsymbol{b}.
\label{eq:psi}
\end{eqnarray}
Here, vectors $\boldsymbol{g},\boldsymbol{h} \in \mathbb{R}^{3N}$ and matrix $\boldsymbol{G} \in \mathbb{R}^{3N} \times\mathbb{R}^{3N}$ are respectively defined as
\begin{equation}
\begin{array}l
\boldsymbol{g} := \big(\boldsymbol{g}_{(1)},\boldsymbol{g}_{(2)}, \cdots, \boldsymbol{g}_{(N)}\big) \\*[1.5mm]
\boldsymbol{h} := \big(\boldsymbol{h}_{(1)},\boldsymbol{h}_{(2)}, \cdots, \boldsymbol{h}_{(N)}\big)
\end{array}
\quad\text{and}\quad
\boldsymbol{G} := \left( \begin{array}{cccc}
             \boldsymbol{G}_{(11)} & \boldsymbol{G}_{(12)} & \ldots & \boldsymbol{G}_{(1N)} \\
             \boldsymbol{G}_{(21)} & \boldsymbol{G}_{(22)} & \ldots & \boldsymbol{G}_{(2N)} \\
             \vdots          & \vdots          & \ddots & \vdots          \\
             \boldsymbol{G}_{(N1)} & \boldsymbol{G}_{(N2)} & \ldots & \boldsymbol{G}_{(NN)}
\end{array} \right),
\end{equation}
whose entries are given by
\begin{equation}
\begin{array}{l}
\boldsymbol{g}_{(i)} :=  (g_{1(i)},g_{2(i)}, g_{3(i)})  \\*[1.5mm]
\boldsymbol{h}_{(i)} :=  (h_{1(i)},h_{2(i)}, h_{3(i)})
\end{array}
\quad\text{and}\quad
\boldsymbol{G}_{(ij)} := \left( \begin{array}{ccc}
             {G}_{11(ij)} & {G}_{12(ij)} & {G}_{13(ij)} \\
             {G}_{21(ij)} & {G}_{22(ij)} & {G}_{23(ij)} \\
             {G}_{31(ij)} & {G}_{32(ij)} & {G}_{33(ij)}
\end{array} \right),
\end{equation}
where
\begin{align*}
&{g}_{1(i)} := \int_{\GM} \Re \left\{\boldsymbol{p}^{11}\si \cdot \overline{(\boldsymbol{u} - \boldsymbol{u}^\ast)} \right\}, \quad
{g}_{2(i)}  := \int_{\GM} \Re \left\{\boldsymbol{p}^{22}\si \cdot \overline{(\boldsymbol{u} - \boldsymbol{u}^\ast)} \right\},  \\
&{g}_{3(i)} := \int_{\GM} \Re \left\{\big(\boldsymbol{p}^{12}\si + \boldsymbol{p}^{21}\si\big) \cdot \overline{(\boldsymbol{u} - \boldsymbol{u}^\ast)} \right\},
\end{align*}
\begin{align*}
&{h}_{1(i)} := \int_{\GM} \Im \left\{-\boldsymbol{p}^{11}\si \cdot \overline{(\boldsymbol{u} - \boldsymbol{u}^\ast)} \right\}, \quad
{h}_{2(i)}  := \int_{\GM} \Im \left\{-\boldsymbol{p}^{22}\si \cdot \overline{(\boldsymbol{u} - \boldsymbol{u}^\ast)} \right\},  \\
&{h}_{3(i)} := \int_{\GM} \Im \left\{-\big(\boldsymbol{p}^{12}\si + \boldsymbol{p}^{21}\si\big) \cdot \overline{(\boldsymbol{u} - \boldsymbol{u}^\ast)} \right\},
\end{align*}
and
\begin{align*}
&{G}_{11(ij)}:= \int_{\GM} \Re \left\{\boldsymbol{p}^{11}\si \cdot \overline{\boldsymbol{p}^{11}\sj} \right\}, \;
{G}_{12(ij)} := \int_{\GM} \Re \left\{\boldsymbol{p}^{11}\si \cdot \overline{\boldsymbol{p}^{22}\sj} \right\}, \;
{G}_{13(ij)} := \int_{\GM} \Re \left\{\boldsymbol{p}^{11}\si \cdot \big(\overline{\boldsymbol{p}^{12}\sj + \boldsymbol{p}^{21}\sj}\big) \right\}, \\
&{G}_{21(ij)}:= \int_{\GM} \Re \left\{\boldsymbol{p}^{22}\si \cdot \overline{\boldsymbol{p}^{11}\sj} \right\}, \;
{G}_{22(ij)} := \int_{\GM} \Re \left\{\boldsymbol{p}^{22}\si \cdot \overline{\boldsymbol{p}^{22}\sj} \right\}, \;
{G}_{23(ij)} := \int_{\GM} \Re \left\{\boldsymbol{p}^{22}\si \cdot \big(\overline{\boldsymbol{p}^{12}\sj + \boldsymbol{p}^{21}\sj}\big) \right\}, \\
&{G}_{31(ij)}:= \int_{\GM} \Re \left\{\big(\boldsymbol{p}^{12}\si + \boldsymbol{p}^{21}\si\big) \cdot \overline{\boldsymbol{p}^{11}\sj} \right\}, \;
{G}_{32(ij)} := \int_{\GM} \Re \left\{\big(\boldsymbol{p}^{12}\si + \boldsymbol{p}^{21}\si\big) \cdot \overline{\boldsymbol{p}^{22}\sj} \right\}, \\
&{G}_{33(ij)}:= \int_{\GM} \Re \left\{\big(\boldsymbol{p}^{12}\si + \boldsymbol{p}^{21}\si\big) \cdot \big(\overline{\boldsymbol{p}^{12}\sj + \boldsymbol{p}^{21}\sj}\big) \right\}. 
\end{align*}

% --------------------------------------------------------------------------
\section{Reconstruction Algorithm} \label{sec:algorithm}
% --------------------------------------------------------------------------

\noindent For each fixed pair $(N,\boldsymbol{z})$, we seek  $(\boldsymbol{a},\boldsymbol{b})$ that minimizes $\Psi$ according to \eqref{eq:psi}. Since $\Psi$ represents a quadratic form with respect to $\boldsymbol{a}$ and $\boldsymbol{b}$, sufficient optimality conditions
\begin{eqnarray}
D_{\boldsymbol{a}} \Psi(N,\boldsymbol{z},\boldsymbol{a},\boldsymbol{b}) \cdot \delta\boldsymbol{a}
&\!\!=\!\!& 0, \quad~ \forall \, \delta\boldsymbol{a} \in \mathbb{R}^{3N},\\
D_{\boldsymbol{b}} \Psi(N,\boldsymbol{z},\boldsymbol{a},\boldsymbol{b}) \cdot \delta\boldsymbol{b}
&\!\!=\!\!&  0, \quad~ \forall \, \delta\boldsymbol{b} \in \mathbb{R}^{3N},
\end{eqnarray}
lead to the linear systems
\begin{equation}
\boldsymbol{G} \boldsymbol{a} = -\boldsymbol{g} \quad\text{and}\quad \boldsymbol{G} \boldsymbol{b} = -\boldsymbol{h}.
\label{eq:system}
\end{equation}
In this setting, the solution $(\boldsymbol{a},\boldsymbol{b})$ of \eqref{eq:system} is implicitly a function of the vector \eqref{bzdef} of source locations $\boldsymbol{z}$, namely $\boldsymbol{a} = \boldsymbol{a}(\boldsymbol{z})$ and $\boldsymbol{b} = \boldsymbol{b}(\boldsymbol{z})$. On substituting \eqref{eq:system} into \eqref{eq:psi}, the optimal vector of source locations $\boldsymbol{z}^{\star}$ can be trivially obtained via combinatorial search over a prescribed grid, $\boldsymbol{Z}$, of $M \geqslant N$ trial source locations geared toward solving the minimization problem
\begin{equation} 
\boldsymbol{z}^{\star} = \underset{\boldsymbol{z} \subset \boldsymbol{Z}}{\text{argmin}}
\left\{ \Psi(N,\boldsymbol{z},\boldsymbol{a}(\boldsymbol{z}),\boldsymbol{b}(\boldsymbol{z})) =
\dfrac{1}{2} \big(\boldsymbol{g} \cdot \boldsymbol{a}(\boldsymbol{z}) + \boldsymbol{h} \cdot \boldsymbol{b}(\boldsymbol{z})\big) \right\}.
\label{Psi}
\end{equation}
On resolving $\boldsymbol{z}^{\star}$, the components of $N$ reconstructed moment tensors $\boldsymbol{M}\si^\star$ are then given by the optimal ``strength'' vectors $\boldsymbol{a}^{\star} = \boldsymbol{a}(\boldsymbol{z}^{\star})$ and $\boldsymbol{b}^{\star} = \boldsymbol{b}(\boldsymbol{z}^{\star})$. The associated optimal value of the objective function is denoted as $\Psi^\star:= \Psi(N,\boldsymbol{z}^\star,\boldsymbol{a}^\star,\boldsymbol{b}^\star)$. We remark that when the ``true" number of micro-seismic sources, $N^*$, is less than~$N$, numerical simulations show that $N-N^*$ pairs $(\boldsymbol{\alpha}_{(i)}^\star,\boldsymbol{\beta}_{(i)}^\star)$ in the solution set $(\boldsymbol{a}^{\star},\boldsymbol{b}^{\star})$ take \emph{near-trivial} values.

To complete the analysis, we next introduce a second-order optimization algorithm that synthesizes the process of obtaining $\boldsymbol{z}^\star$ and $(\boldsymbol{a}^{\star},\boldsymbol{b}^{\star})$ from the computational point of view. The input of the algorithm is listed below:
\begin{itemize}
\item Upper bound $N$ on the number of (micro-seismic) point sources.
\item Grid $\boldsymbol{Z}$ of $M \geqslant N$ trial source locations.
\item Canonical solutions $\boldsymbol{p}^{kl}\si$ for each grid point $\boldsymbol{\xi}\si\in\boldsymbol{Z}$.
\end{itemize}
The algorithm returns the optimal set of source locations $\boldsymbol{z}^{\star}$ and respective moment tensor components given by $(\boldsymbol{a}^{\star},\boldsymbol{b}^{\star})$. The above procedure, originally developed in \cite{CanelasJCP2014} in the context of inverse potential problems, is shown in Algorithm~\ref{alg:invprob} using pseudo-code format. Therein, $\Pi : \{1,2,\ldots,M\}^{N} \mapsto \boldsymbol{Z}$ maps the vector of source indices $\mathcal{I} = (i_{1}, i_{2},\ldots, i_{N})$ to the corresponding vector of source locations $\boldsymbol{z} \subset \boldsymbol{Z}$. For further applications of this algorithm, we refer to \cite{NovotnyJOTA2019c}.

\IncMargin{1em}
\begin{algorithm}[!t]
\SetAlgoLined
\SetKwInOut{Input}{input}\SetKwInOut{Output}{output}
\LinesNumbered
\Input{$N$, $\boldsymbol{Z}$, $\boldsymbol{p}^{kl}\si$ $\forall \boldsymbol{\xi}\si\!\in\boldsymbol{Z}$}
\BlankLine
Initialization:\: $\boldsymbol{z}^{\star}\leftarrow \boldsymbol{0}$;\quad $(\boldsymbol{a}^{\star},\boldsymbol{b}^{\star}) \leftarrow (\boldsymbol{0},\boldsymbol{0})$;\quad $\Psi^{\star} \leftarrow \infty$;\quad $M \leftarrow \text{card}(\boldsymbol{Z})$ \\
\For{$i_{1} \leftarrow 1$ \KwTo $M$}
{
	\For{$i_{2} \leftarrow i_{1}+1$ \KwTo $M$}
	{
		\vdots
		\For{$i_{N} \leftarrow i_{N-1}+1$ \KwTo $M$}
		{
            $\boldsymbol{g} \leftarrow 	
            \begin{bmatrix}
			\boldsymbol{g}_{(i_{1})} \\
  			\boldsymbol{g}_{(i_{2})} \\
  			\vdots       \\
  			\boldsymbol{g}_{(i_{N})} \\
  			\end{bmatrix}$; \quad	
  			$\boldsymbol{h} \leftarrow 	
            \begin{bmatrix}
			\boldsymbol{h}_{(i_{1})} \\
  			\boldsymbol{h}_{(i_{2})} \\
  			\vdots       \\
  			\boldsymbol{h}_{(i_{N})} \\
  			\end{bmatrix}$; \quad	
			$\boldsymbol{G} \leftarrow
  			\begin{bmatrix}
			\boldsymbol{G}_{(i_{1}i_{1})} & \boldsymbol{G}_{(i_{1}i_{2})}  & \cdots & \boldsymbol{G}_{(i_{1}i_{N})}  \\
			\boldsymbol{G}_{(i_{2}i_{1})} & \boldsymbol{G}_{(i_{2}i_{2})}  & \cdots & \boldsymbol{G}_{(i_{2}i_{N})}  \\
  			\vdots & \vdots & \ddots & \vdots\\
			\boldsymbol{G}_{(i_{N}i_{1})} & \boldsymbol{G}_{(i_{N}i_{2})}  & \cdots & \boldsymbol{G}_{(i_{N}i_{N})}  \\
  			\end{bmatrix}$ \\*[1mm]
            $\boldsymbol{a} \leftarrow - \boldsymbol{G}^{-1}\boldsymbol{g}$;\quad 
            $\boldsymbol{b} \leftarrow - \boldsymbol{G}^{-1}\boldsymbol{h}$;\quad
            $\Psi\leftarrow \displaystyle\tfrac{1}{2}(\boldsymbol{g}\cdot\boldsymbol{a}+\boldsymbol{h}\cdot\boldsymbol{b})$ \\
            $\mathcal{I} \leftarrow (i_{1}, i_{2},\ldots, i_{N})$;\quad $\boldsymbol{z} \leftarrow \Pi(\mathcal{I})$ \\

 	   		\If{$\Psi< \Psi^{\star}$}
	    		{
                $\boldsymbol{z}^{\star}\leftarrow \boldsymbol{z}$;\quad $(\boldsymbol{a}^{\star},\boldsymbol{b}^{\star})\leftarrow (\boldsymbol{a},\boldsymbol{b})$;\quad $\Psi^{\star}\leftarrow \Psi$ \\
 	   			}
		}
	}
}
\Return{$\boldsymbol{z}^{\star}$, $(\boldsymbol{a}^{\star},\boldsymbol{b}^{\star})$, $\Psi^{\star}$}
\caption{Micro-seismic source reconstruction}
\label{alg:invprob}
\end{algorithm}

{\color{black}
In  Algorithm \ref{alg:invprob}, optimal source locations $\boldsymbol{z}^{\star}$ are obtained through a combinatorial search over $M$ trial points sampling the set of admissible locations $\boldsymbol{Z}$. As a result, the computational complexity $\mathcal{C}(M,N)$ of the algorithm can be evaluated by the formula
\begin{equation*}
\mathcal{C}(M,N) \;\approx\;
\left(
\begin{array}{c}
M \\
N
\end{array}
\right)
N^{3}
\;=\; \displaystyle\frac{M!}{N!(M - N)!}N^{3}.
\label{eq:complexity}
\end{equation*}
In Fig.~\ref{fig:complexity}, the graphs of $N \times \log_{10}(\mathcal{C}(M,N))$ for $M=100$ and $M=400$ are plotted as solid and dashed lines, respectively. As can be seen from the display, the computational cost of the algorithm may become prohibitive for $N \approx M/2$. In the ensuing numerical examples (Section \ref{sec:umerics}), we set $N \ll M$, so that Algorithm \ref{alg:invprob} runs in a few seconds for all examples.

\begin{figure}[!t]
\begin{center}
\includegraphics[scale = 1.3]{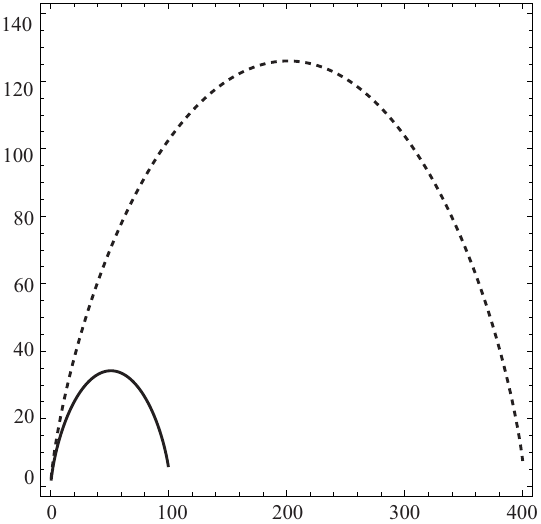}
\end{center}
\caption{Complexity order of Algorithm \ref{alg:invprob}: $N\times \log_{10}(\mathcal{C}(M,N))$ for $M=100$ (solid) and $M=400$ (dashed).}
\label{fig:complexity}
\end{figure}}

\begin{remark}
In the standard (time-domain) interpretation of acoustic emission signals~\citep{Scru1985}, the unknown onset ``$t=0$'' of a micro-seismic event, see~\eqref{zerotime}, requires the analysis to be reformulated in terms of relative arrival times -- which results in a nonlinear minimization problem. In the context of~\eqref{Psi}, on the other hand, we find by the translation property 
\[
\mathcal{F}[g(t+\Delta t)](\omega) \;=\; e^{\mathfrak{i}\omega \Delta t} \, \mathcal{F}[g(t)](\omega)  
\]
of the Fourier transform that an unknown onset, $\Delta t\si$, of the ``(i)''th micro-seismic event (relative to~$t=0$ implicit to the Fourier transform) affects only the phase of $\boldsymbol{M}\si^\star=\boldsymbol{M}\si^\star(\omega)$ via factor~$e^{\mathfrak{i}\omega \Delta t\si}$. As a result, we see that Algorithm~\ref{alg:invprob} yields the event locations~$\xi\si$ and moduli, $|\boldsymbol{M}\si^\star|$, of the respective moment tensors that are invariant with respect to the unknown onsets~$\Delta t\si$. To highlight the performance of the frequency-domain scheme, we implicitly assume~$\Delta t\si=0$ in the ensuing examples. 
\end{remark}

% --------------------------------------------------------------------------
\section{Numerical Results} \label{sec:umerics}
% --------------------------------------------------------------------------

\noindent {\color{black}Thanks to the fact that the moment tensor $\boldsymbol{M}\si\in\mathbb{C}^{2\times 2}$ is symmetric, its eigenvalues can be conveniently written as
\begin{equation}
m\si^{1,2} := \frac{1}{2} \left(\mathrm{tr} (\boldsymbol{M}\si) \pm \sqrt{\boldsymbol{M}\si^D : \boldsymbol{M}\si^D} \right)\
\end{equation}
in terms of the volumetric $\mathrm{tr} (\boldsymbol{M}\si)$ and deviatoric $\boldsymbol{M}\si^D$ components of~$\boldsymbol{M}\si$, with
\begin{equation}
\boldsymbol{M}\si^D \,=\, \boldsymbol{M}\si - \frac{1}{2} \mathrm{tr} (\boldsymbol{M}\si) \boldsymbol{I}_2.
\end{equation}}
In the sequel, we denote the affiliated eigenvectors by $\boldsymbol{v}\si^{1,2}$. 

\begin{figure}[!ht]
\begin{center}
\subfigure[$\tfrac{m\si^1}{\gamma\si} = \tfrac{m\si^2}{\gamma\si}> 0$]{\includegraphics[scale = 0.95]{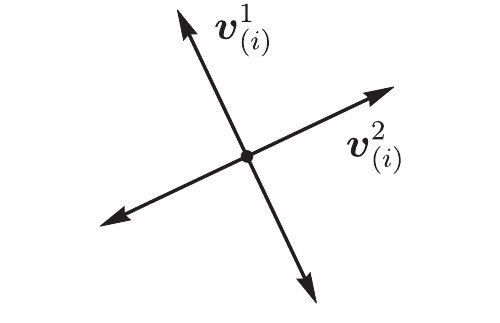} \label{fig:cavita}} 
\subfigure[$\tfrac{m\si^1}{\gamma\si} > \tfrac{m\si^2}{\gamma\si} > 0$]{\includegraphics[scale = 0.95]{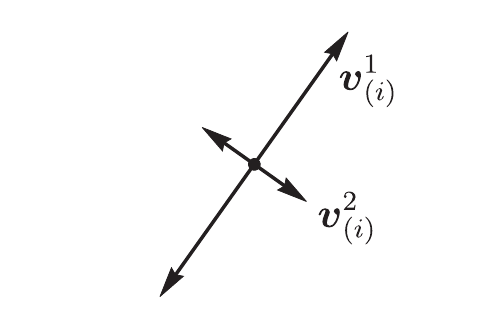} \label{fig:tensile}} 
\subfigure[$\tfrac{m\si^1}{\gamma\si} > 0 > \tfrac{m\si^2}{\gamma\si}$]{\includegraphics[scale = 0.95]{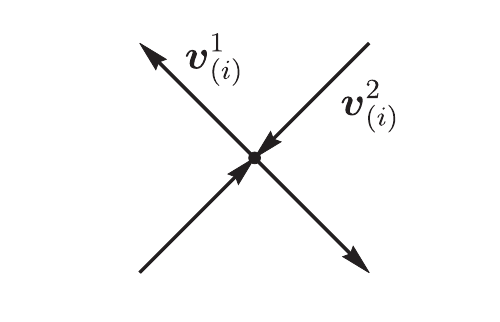} \label{fig:shear}} \vspace*{-3mm}
\end{center}
\caption{Representation of the moment tensors $\boldsymbol{M}\si^*$ in terms of their eigenvalues $m\si^{1,2}$ and eigenvectors $\boldsymbol{v}\si^{1,2}$: (a) cavitation, (b) mode I crack, and (c) mode II crack.} 
\label{fig2}
\end{figure}

For the purposes of source inversion, we next consider three types of micro-seismic events given by the moment tensors $\boldsymbol{M}^*\si\in\mathbb{C}^{2\times 2}$ ($i=\overline{1,N^*}$) featuring: (i) complex amplitude $\gamma\si \in \mathbb{C}$, (ii) unit normal to the microcrack $\boldsymbol{\eta}\si\in\mathbb{R}^2$ (when applicable), and (iii)  Lam\'{e} moduli $\mu$ and $\lambda$ of the background solid~\citep{Aki2002}. Specifically, when generating the synthetic data $\boldsymbol{u}^*$ according to~\eqref{eq:strongZ} and~\eqref{truef}, we allow for  
\begin{enumerate}
    \item Cavitation: 
    \begin{equation}
        \boldsymbol{M}\si^* = 2 \gamma\si (\mu + \lambda) \boldsymbol{I}_2 \quad \Rightarrow \quad m\si^{1,2} = 2 \gamma\si (\mu + \lambda);
        \label{eq:cavita}
    \end{equation}
    \item Mode I crack: 
    \begin{equation}
        \boldsymbol{M}\si^* = \gamma\si (2 \mu (\boldsymbol{\eta}\si \otimes \boldsymbol{\eta}\si) + \lambda \boldsymbol{I}_2) \quad \Rightarrow \quad m\si^1 = \gamma\si (2 \mu + \lambda), ~ m\si^2 = \gamma\si \lambda;
        \label{eq:tensile}
    \end{equation}
    \item Mode II crack: 
    \begin{equation}
        \boldsymbol{M}\si^* = \gamma\si \mu (\boldsymbol{\eta}\si^{\perp} \otimes \boldsymbol{\eta}\si + \boldsymbol{\eta}\si \otimes \boldsymbol{\eta}\si^{\perp}) \quad \Rightarrow \quad m\si^{1,2} = \pm \gamma\si \mu.
        \label{eq:shear}
    \end{equation}    
\end{enumerate}
For future reference, the moment tensors given by~\eqref{eq:cavita}--\eqref{eq:shear} are depicted graphically in Fig.~\ref{fig2}.

\subsection{Testing setup}

\begin{figure}[htbp]
\begin{center}
\includegraphics[scale = 0.3]{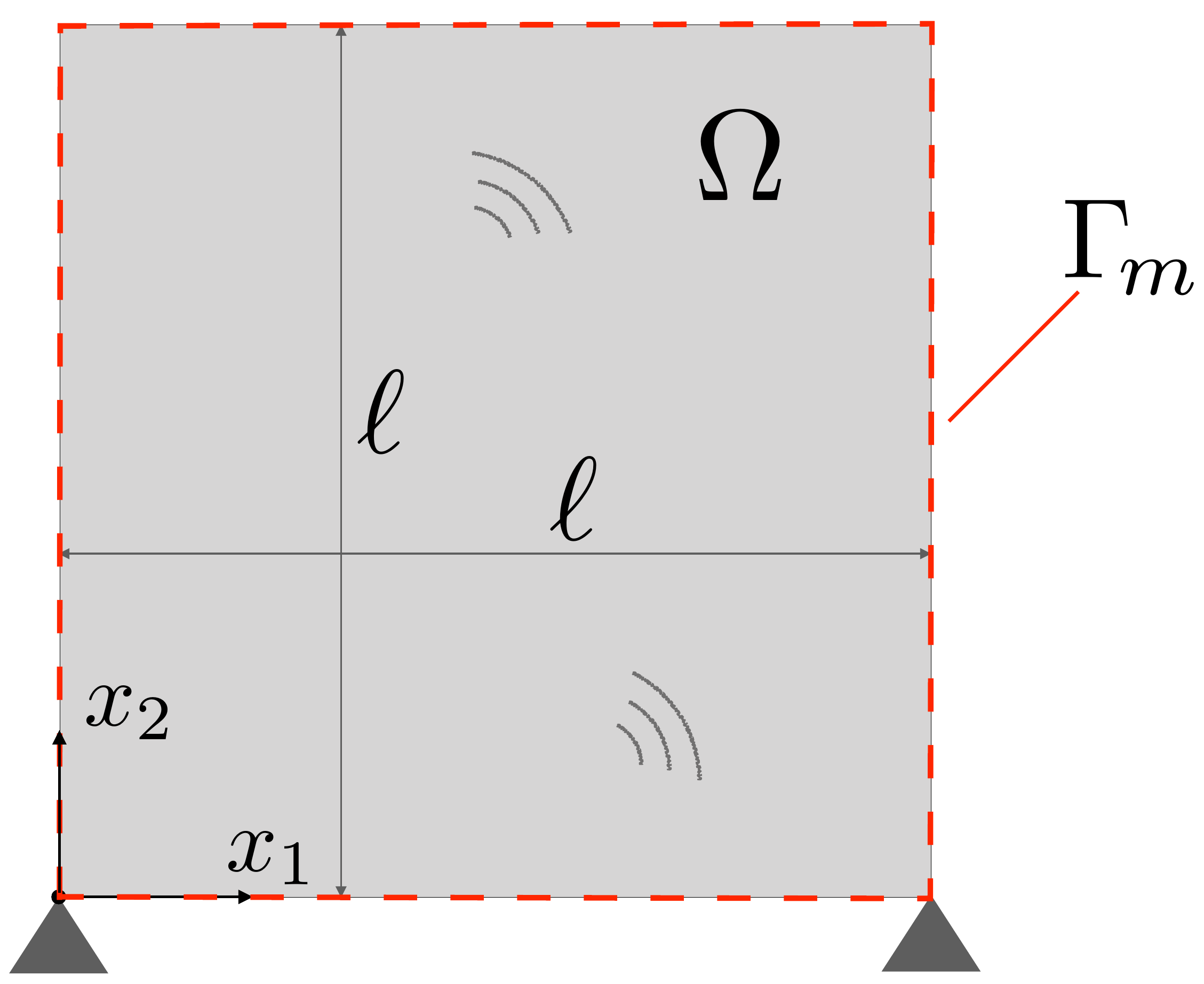} \vspace*{-3mm}
\end{center}
\caption{Square ``rock'' specimen undergoing acoustic emission.}
\label{fig:exex}
\end{figure}

\noindent The elastic body~$\Omega$ used for numerical simulations is taken as an $\ell\times\ell$ block of ``rock'' with mass density $\rho$ and Lam\'{e} moduli $\lambda=\mu$ (Poisson's ratio $\nu=0.25$), fixed at the bottom corners as in Fig.~\ref{fig:exex}. The pointwise motion sensors are assumed to be distributed along the boundary $\partial\Omega$ with various densities and apertures as described in the sequel. The dimensionless frequency of acoustic emission is taken as
\[
\bar{\omega} \,=\, \frac{\omega \,\ell}{\sqrt{\mu/\rho}} \,=\, 10\pi,
\]
resulting in the specimen-size-to-shear-wavelength ratio of $\ell/\lambda_s=5$. With reference to~\eqref{mtensor}, \eqref{eq:fdelta} and~\eqref{eq:cavita}--\eqref{eq:shear}, we also introduce  the dimensionless coordinates $\bar{\boldsymbol{x}}=\ell^{-1}\boldsymbol{x}$; we consider the dimensionless source strength $\bar{\gamma} = \ell^{-3}\gamma$, and we specify the unit normal to the microcrack as $\boldsymbol{\eta} = (\cos \theta, \sin \theta)$, where $\theta$ is the angle measured counter-clockwise from the horizontal axis. The forward elastodynamic problem is solved via standard Galerkin finite element method. To handle the germane wave propagation with sufficient accuracy, domain $\Omega$ is first subdivided into a uniform  $10 \times 10$ grid of square subdomains. Then, each subdomain is discretized via $4^{n}$ triangular finite elements with $n = 7$. Next, the set of admissible source locations $\boldsymbol{Z}$ is taken as the union of vertices of like triangles with $n = 1$, giving $M=221$ in Algorithm~\ref{alg:invprob}. To illustrate the performance of the inversion algorithm, we adopt the graphical representation of moment tensors introduced in Fig.~\ref{fig2}, and we denote the ``true'' (resp. reconstructed) sources by thick red (resp. thin blue) arrows.

In the sequel we tackle several test problems, dealing with both isolated and co-existing sources of acoustic emission. We first consider an idealized scenario where the locations of microcracks belong to the set of admissible locations $\boldsymbol{Z}$, and then proceed to the reconstruction of arbitrarily-located sources.

\begin{remark}
\noindent \textcolor{black}{In what follows, our target application is the acoustic emission (AE) analysis of failure processes in quasi-brittle laboratory samples. Depending on the loading mechanism, either majority of the specimen's surface (e.g.~non-uniform thermal expansion or drying shrinkage), a good part of the surface of the specimen (e.g. split cylinder testing), or only its ``sides'' (e.g. uniaxial compression) may be available for AE sensing. In this vein, our numerical studies assume square specimen geometry and cover situations where the part of the external surface that is available for AE sensing entails anywhere from one to four sides of the square.}
\end{remark}

%--------------------------------------------------------------------------
\subsection{Single cavitation event ($\boldsymbol{\xi}\ssi^*\!\in\boldsymbol{Z}$)}\label{1}
%--------------------------------------------------------------------------

\begin{figure}[htbp]
\begin{center}
\subfigure[Real part]{\includegraphics[scale = 0.8]{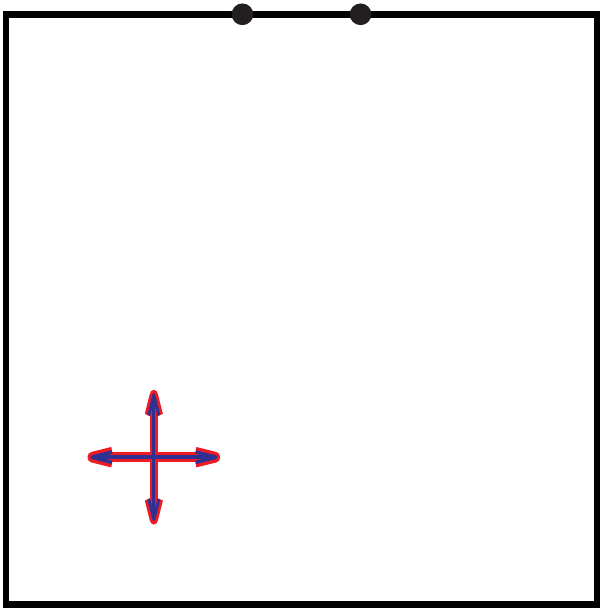}} \qquad
\subfigure[Imaginary part]{\includegraphics[scale = 0.8]{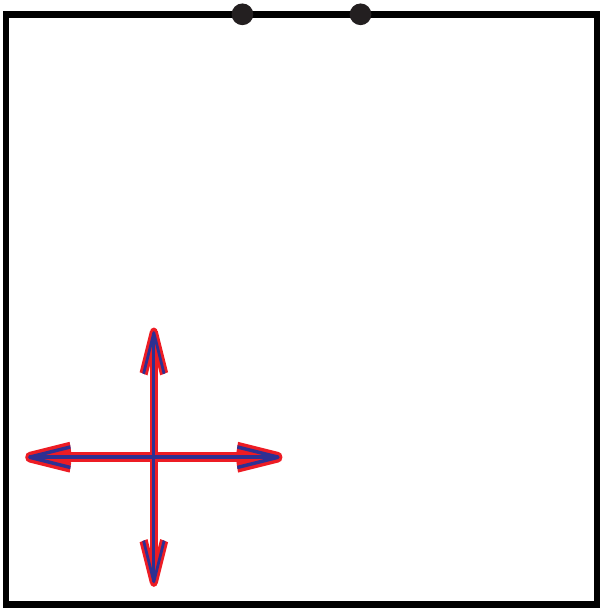}}
\end{center} \vspace*{-3mm}
\caption{Reconstruction of a single micro-seismic source using two biaxial motion sensors.} 
\label{fig:exe1}
\end{figure}

\noindent In the first example we aim to reconstruct a single micro-seismic source of type~\eqref{eq:cavita}, with complex amplitude $\bar{\gamma}_{(1)} = 0.01 + 0.02 \mathfrak{i}$ and location $\boldsymbol{\xi}\ssi^*\!\in\boldsymbol{Z}$, by using a pair of biaxial motion sensors placed on the top surface of the specimen. Table~\ref{tab:ex1:a} lists the respective coordinates of the source and motion sensors. As expected, the source reconstruction shown in Fig.~\ref{fig:exe1} is practically exact.

\begin{table}[htbp]
\centering
\caption{Source and sensor locations for the single event example. \label{tab:ex1:a}} \vspace*{2pt} 
\begin{tabu}{ccc}
\tabucline[1 pt]{-} \\*[-11pt]
 Source   & Sensor &  $\bar{\boldsymbol{\xi}}\si^*$ or $\bar{\boldsymbol{x}}$ \\ \tabucline[1 pt]{-}
Cavitation & &  $(0.25, 0.25)$   \\
&  $\# 1$  &    $(0.40, 1.00)$   \\
&  $\# 2$  &    $(0.60, 1.00)$   \\
\tabucline[1 pt]{-}
\end{tabu}
\end{table}

% --------------------------------------------------------------------------
\subsection{Two co-existing events ($\boldsymbol{\xi}\si^*\!\in\boldsymbol{Z}$).}\label{2}
% --------------------------------------------------------------------------

\begin{figure}[htbp]
\begin{center}
\subfigure[Real part]{\includegraphics[scale = 0.8]{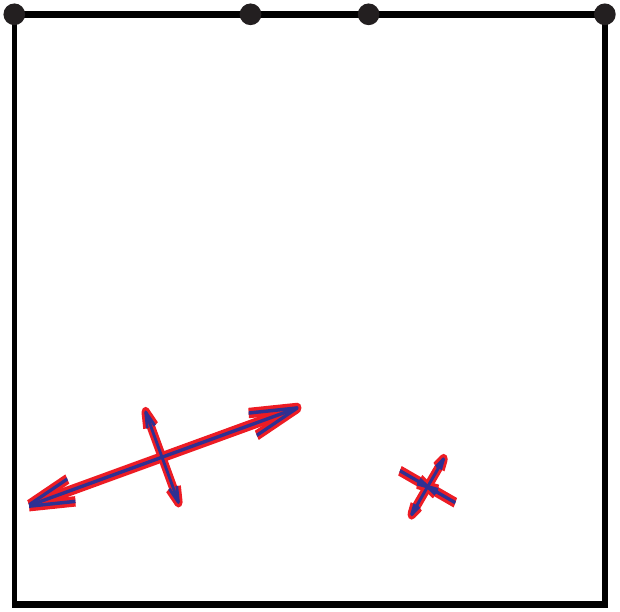}} \qquad
\subfigure[Imaginary part]{\includegraphics[scale = 0.8]{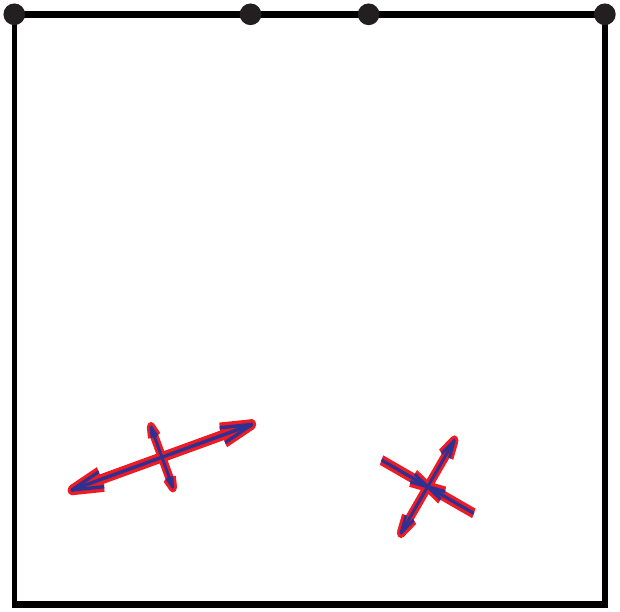}}
\end{center} \vspace*{-3mm}
\caption{Reconstruction of a pair of micro-seismic sources using two biaxial motion sensors.}
\label{fig:exe2}
\end{figure}

\noindent We next seek to reconstruct two micro-seismic sources representing: (i) mode I crack with $\bar{\gamma}_{(1)} = 0.05 + 0.03 \mathfrak{i}$ and $\theta_{(1)} = 20^{\circ}$, and (ii) mode II crack with $\bar{\gamma}_{(2)} = 0.03 + 0.05 \mathfrak{i}$ and $\theta_{(2)} = 15^{\circ} $. As before, we make use of two sensors located on the top surface of the specimen. Table \ref{tab:ex1:b} lists the source and sensor coordinates, the former being limited to the set of admissible locations~$\boldsymbol{Z}$. Again, the reconstruction is nearly exact as shown in Fig. \ref{fig:exe2}.

\begin{table}[htbp]
\centering
\caption{Source and sensor locations for the dual event example. \label{tab:ex1:b}} \vspace*{4pt} 
\begin{tabu}{ccc}
\tabucline[1 pt]{-} \\*[-11pt]
 Source   & Sensor &  $\bar{\boldsymbol{\xi}}\si^*$ or $\bar{\boldsymbol{x}}$  \\  \tabucline[1 pt]{-}
Mode I crack     & &   $(0.20, 0.20)$   \\
Mode II crack    & &   $(0.70, 0.20)$   \\
&  $\# 1$   &     $(0.00, 1.00)$   \\
&  $\# 2$   &     $(0.60, 1.00)$   \\
&  $\# 3$   &     $(0.40, 1.00)$   \\
&  $\# 4$   &     $(1.00, 1.00)$   \\
\tabucline[1 pt]{-}
\end{tabu}
\end{table}

% --------------------------------------------------------------------------
\subsection{Three co-existing events  ($\boldsymbol{\xi}\si^*\!\in\boldsymbol{Z}$).}\label{3}
% --------------------------------------------------------------------------

\begin{figure}[htbp]
\begin{center}
\subfigure[Real part]{\includegraphics[scale = 0.8]{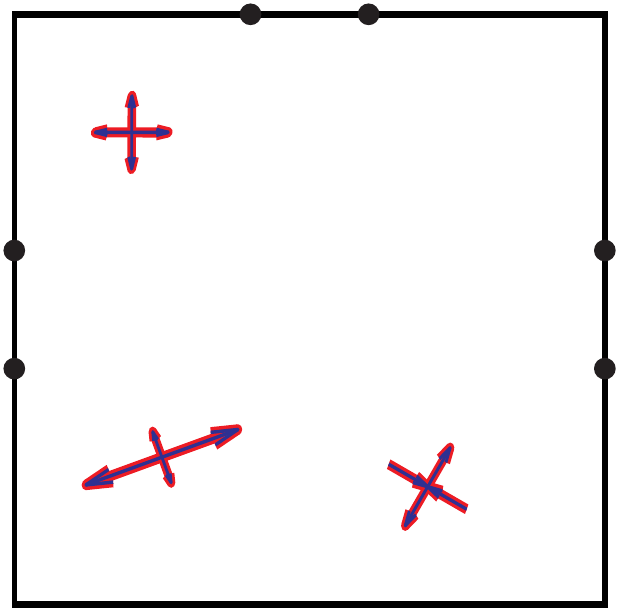}} \qquad
\subfigure[Imaginary part]{\includegraphics[scale = 0.8]{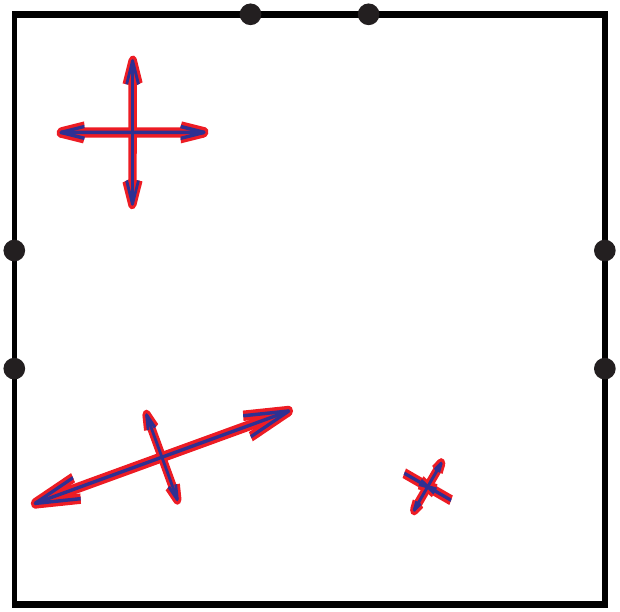}}
\end{center} \vspace*{-3mm}
\caption{Reconstruction of a triplet of micro-seismic sources using six biaxial motion sensors.}
\label{fig:exe3}
\end{figure}

\noindent In this example, we pursue reconstruction of three micro-seismic sources representing: (i) mode I crack with $\bar{\gamma}_{(1)} = 0.03 + 0.05 \mathfrak{i}$ and $\theta_{(1)} = 20^{\circ} $; (ii) mode II crack with $\bar{\gamma}_{(2)} = 0.05 + 0.03 \mathfrak{i}$ and $\theta_{(2)} = 15^{\circ}$, and (iii) cavitation with $\bar{\gamma}_{(3)} = 0.01 + 0.02 \mathfrak{i}$. As sensory data, we consider the biaxial motion measurements captured by three pairs of sensors shown in Fig.~\ref{fig:exe3}. For completeness, Table \ref{tab:ex1:c} lists the featured source and sensor coordinates, the former being limited to the set of admissible locations~$\boldsymbol{Z}$. As can be seen from Fig.~\ref{fig:exe3}, the quality of triple source reconstruction is commensurate with that in previous examples.

\begin{table}[htbp]
\centering
\caption{Source and sensor locations for the triple event example. \label{tab:ex1:c}}  
\begin{tabu}{ccc}
\tabucline[1 pt]{-} \\*[-11pt]
 Source   & Sensor &  $\bar{\boldsymbol{\xi}}\si^*$ or $\bar{\boldsymbol{x}}$  \\  \tabucline[1 pt]{-}
Mode I crack  & &   $(0.25, 0.25)$     \\
Mode II crack & &   $(0.70, 0.20)$     \\
Cavitation    & &   {\color{black}$(0.20, 0.80)$}     \\
&  $\# 1$     &     $(0.40, 1.00)$     \\
&  $\# 2$     &     $(0.60, 1.00)$	   \\
&  $\# 3$     &     $(0.00, 0.40)$     \\
&  $\# 4$     &     $(0.00, 0.60)$	   \\
&  $\# 5$     &     $(1.00, 0.40)$     \\
&  $\# 6$     &     $(1.00, 0.60)$	   \\
\tabucline[1 pt]{-}
\end{tabu}
\end{table}

\begin{remark}
At this point, it is worth noting that the reconstruction fails if a smaller-than-featured number of sensors is deployed in each of the foregoing examples. Qualitatively speaking, this suggests the use of at least two sensors per (micro-seismic) source. \textcolor{black}{When using~$M$ sensors in a laboratory setting, one should accordingly expect to reliably reconstruct up to~$M/2$ simultaneous sources. In situations where the reconstruction algorithm consistently exposes~$\geqslant M/2$ contemporaneous events, the above result suggests either (i) deploying additional motion sensors, or (ii) retaining only the "strongest" $M/2$ events, as quantified e.g. in terms of Frobenius norm of the moment tensors~~$\boldsymbol{M}\si$, $i=\overline{1,N}$. For completeness, we note that in conventional acoustic emission (AE) testing~\citep{Gros2008}, micro-seismic events are reconstructed one at a time -- which precludes the existence of contemporaneous sources.}
\end{remark}

% --------------------------------------------------------------------------
\subsection{Two co-existing events ($\boldsymbol{\xi}\si^*\!\notin\boldsymbol{Z}$).}
% --------------------------------------------------------------------------
\noindent We next consider a more realistic scenario where the ``true'' source positions~$\boldsymbol{\xi}\si^*$ do not belong to the set of admissible locations $\boldsymbol{Z}$. The idea is to start with a ``rough'' grid search in terms of $\boldsymbol{Z}$, and to follow up with recursive grid refinement around previously recovered source locations -- up to a prescribed stopping criterion.

In this example, we use 16 biaxial sensors distributed uniformly over~$\partial\Omega$ to reconstruct two co-existing events: (i) mode I crack with $\bar{\gamma}_{(1)} = 0.05 + 0.03 \mathfrak{i}$ and $\theta_{(1)} = 20^{\circ}$, and (ii) mode II crack with $\bar{\gamma}_{(2)} = 0.03 + 0.05 \mathfrak{i}$ and~$\theta_{(2)} = 15^{\circ}$. Table \ref{tab:ex2} specifies the source locations, neither of which belongs to the set of admissible locations~$\boldsymbol{Z}$. For generality, we further assume that the exact number of sources is unknown by setting $N=3>N^*=2$.

\begin{figure}[htbp]
\begin{center}
\subfigure[iteration $\# 1$]{\includegraphics[scale = 0.6]{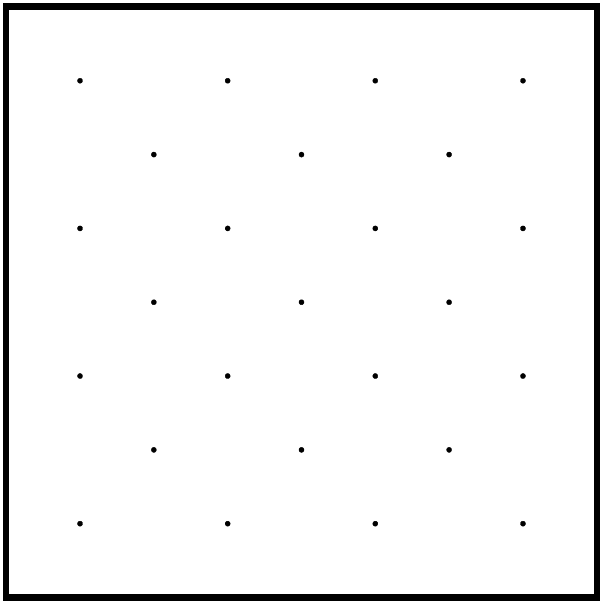}
\label{fig:exe6c:a}} \qquad
\subfigure[iteration $\# 2$]{\includegraphics[scale = 0.6]{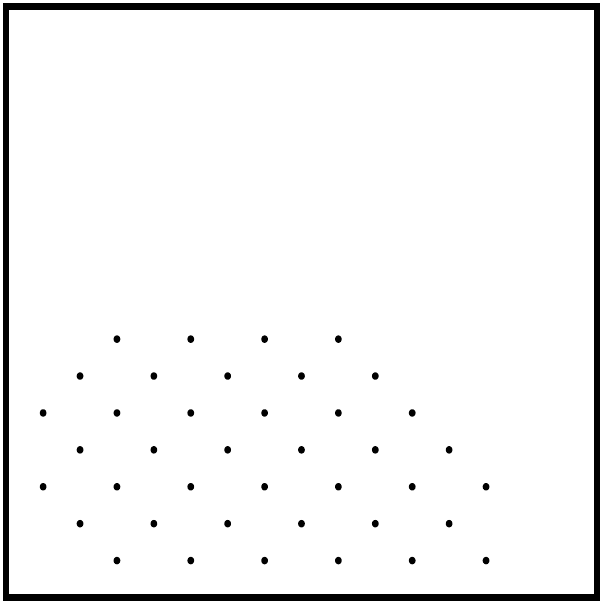}
\label{fig:exe6c:b}} \\
\subfigure[iteration $\# 3$]{\includegraphics[scale = 0.6]{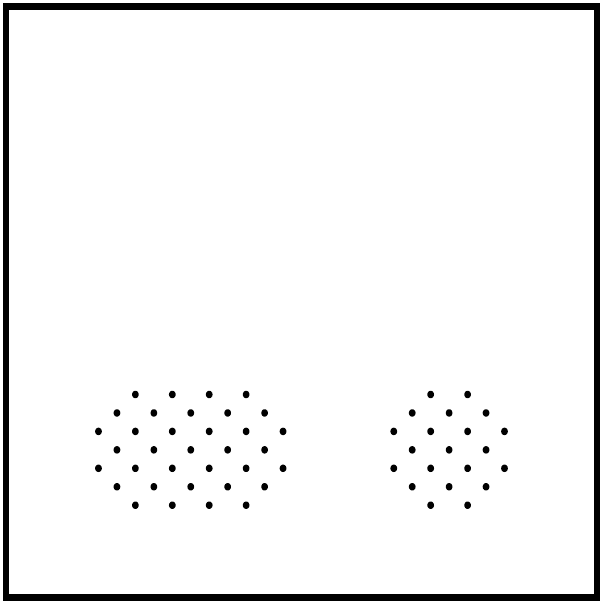}
\label{fig:exe6c:c}} \qquad
\subfigure[iteration $\# 4$]{\includegraphics[scale = 0.6]{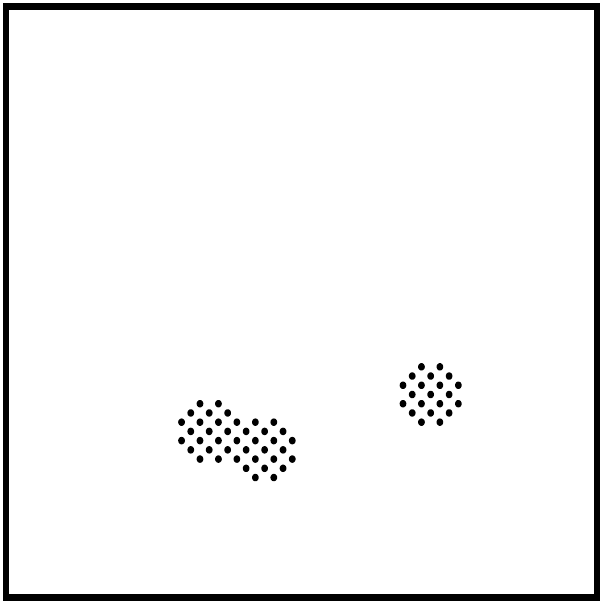}
\label{fig:exe6c:d}}
\end{center} \vspace*{-3mm}
\caption{Grid search refinements $\boldsymbol{Z}_1$ through $\boldsymbol{Z}_4$.}
\label{fig:exe6c}
\end{figure}

To initiate the recursive search algorithm, we first subdivide $\Omega$ into a uniform $4\times 4$ grid of square regions. Then, each $\tfrac{\ell}{4}\times\tfrac{\ell}{4}$ region is further split into $4^{n}$ triangles, using $n = 8$ for the computational mesh and letting $n=1$ to establish the initial set, $\boldsymbol{Z}_1$, of admissible source locations shown in Fig.~\ref{fig:exe6c:a}. Since $\boldsymbol{\xi}\si^* \notin \boldsymbol{Z}_1$, the vector of reconstructed locations $\boldsymbol{z}^{\star}_1$  is found to contain a set of nodes surrounding the exact locations. Next, the set of admissible locations $\boldsymbol{Z}_1$ is replaced by a denser grid, $\boldsymbol{Z}_{2}^\prime$, obtained by letting $n = 2$. Then, a new set of admissible locations $\boldsymbol{Z}_2$ -- shown in Fig. \ref{fig:exe6c:b} -- is constructed as the restriction $\boldsymbol{Z}_{2}^\prime$ to circular regions of radius $\ell/2^{n}$ centred at $\boldsymbol{z}^{\star}_1$. By setting $n \leftarrow n+1$, the process is repeated up to $n = 8$, resulting in eight iterations of adaptive grid refinement. As an illustration, Fig.~\ref{fig:exe6c:c} and Fig.~\ref{fig:exe6c:d} plot respectively the refinements~$\boldsymbol{Z}_3$ and~$\boldsymbol{Z}_4$.

\begin{figure}[htbp]
\begin{center}
\subfigure[Real part]{\includegraphics[scale = 0.8]{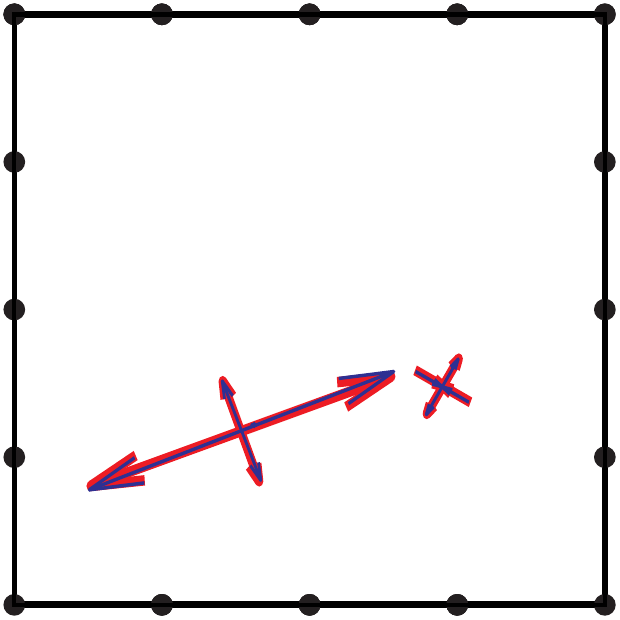}} \qquad
\subfigure[Imaginary part]{\includegraphics[scale = 0.8]{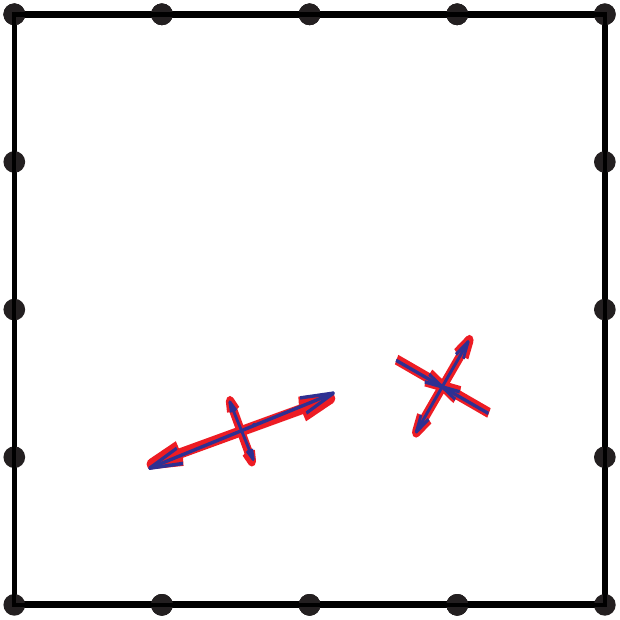}}
\end{center} \vspace*{-3mm}
\caption{Reconstruction of a dual ``off-grid" micro-seismic source using sixteen biaxial motion sensors: iteration $n=7$.}
\label{fig:exe6a7}
\end{figure}

\begin{figure}[htbp]
\begin{center} \vspace*{3mm}
\subfigure[Real part]{\includegraphics[scale = 0.8]{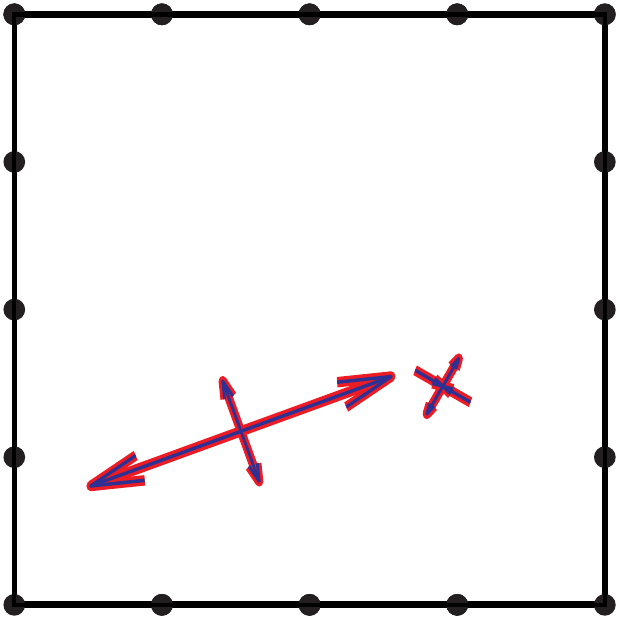}} \qquad
\subfigure[Imaginary part]{\includegraphics[scale = 0.8]{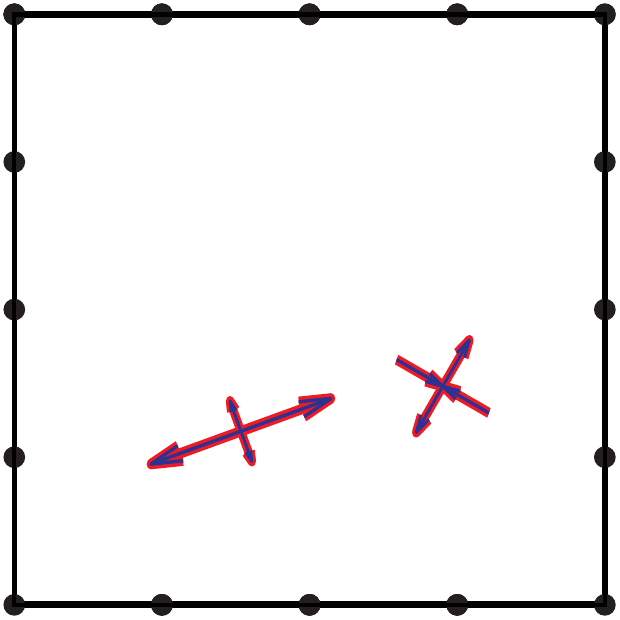}}
\end{center} \vspace*{-3mm}
\caption{Reconstruction of a dual ``off-grid" micro-seismic source using sixteen biaxial motion sensors: iteration $n=8$.}
\label{fig:exe6a8}
\end{figure}

\begin{figure}[htbp]
\begin{center} \vspace*{3mm}
\includegraphics[scale = 0.85]{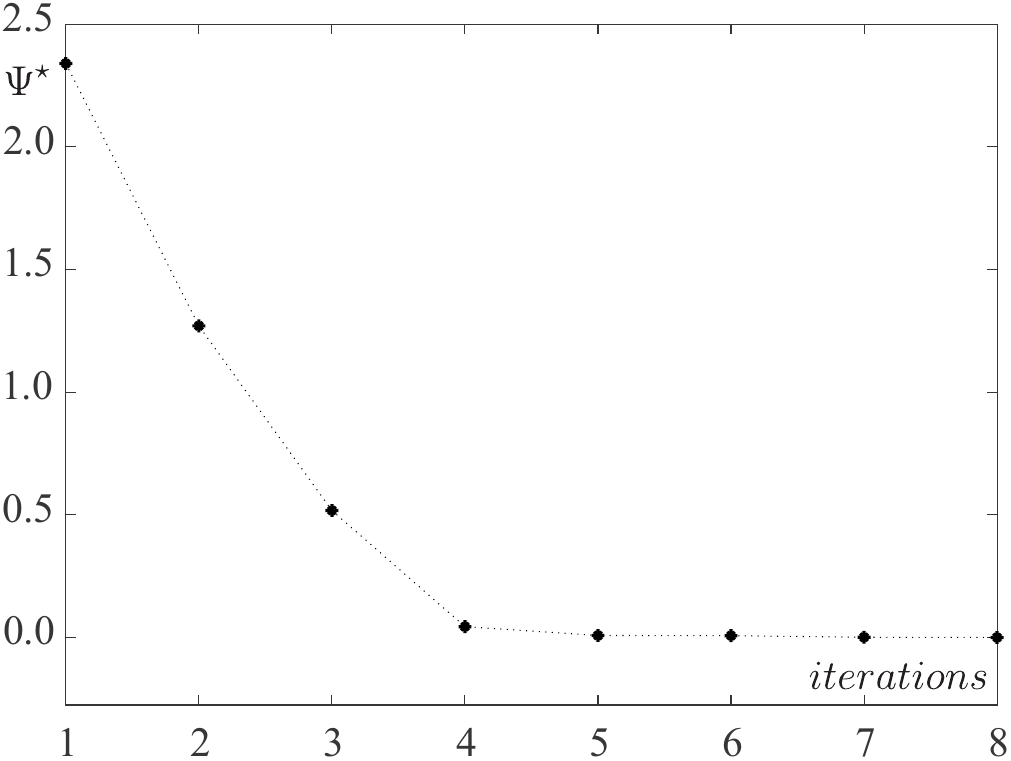}
\end{center} \vspace*{-3mm}
\caption{Variation of the objective functional $\Psi^\star$ during adaptive grid refinement.}
\label{fig:exe6b}
\end{figure}

The source reconstructions given by the last two iterations ($n=7$ and $n=8$) are shown respectively in Fig.~\ref{fig:exe6a7} and Fig.~\ref{fig:exe6a8}. In each case, the two events are well resolved in terms of both location and moment tensor. Due to the  premise~$N=3$, a third fault is also found, but with a negligible strength (invisible in the diagrams). Note that $\boldsymbol{\xi}\si^* \notin \boldsymbol{Z}_7$ but $\boldsymbol{\xi}\si^* \in \boldsymbol{Z}_8$, which explains nearly exact reconstruction obtained for $n=8$ and a small distortion observed for $n=7$. For completeness, diminishing values of the cost functional~$\Psi^\star$ stemming from~\eqref{Psi} during the iterative reconstruction process are shown in Fig.~\ref{fig:exe6b}.

\begin{table}[htbp]
\centering
\caption{Source locations for the dual ``off-grid'' event example. \label{tab:ex2}}
\begin{tabu}{cc}
\tabucline[1 pt]{-} \\*[-11pt]
 Source   & $\bar{\boldsymbol{\xi}}\si^*$  \\  \tabucline[1 pt]{-}
Mode I crack    &  $(0.3837, 0.2939)$   \\
Mode II crack   &  $(0.7257, 0.3700)$   \\
\tabucline[1 pt]{-}
\end{tabu}
\end{table}

\subsection{Reconstruction under random modeling errors ($\boldsymbol{\xi}\si^*\!\in\boldsymbol{Z}$).}

\begin{figure}[htbp]
\begin{center} 
\includegraphics[scale = 0.9]{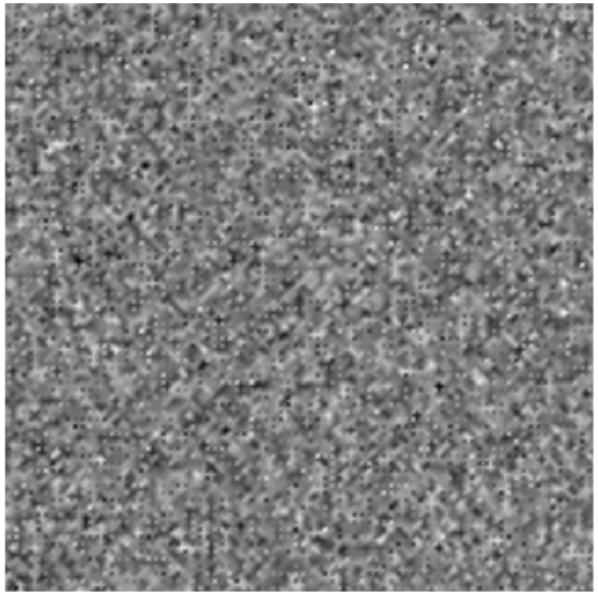}
\end{center}\vspace*{-3mm}
\caption{Spatial variation of the multiplier $(1+\eta\tau)$ used to perturb the background material properties ($\eta=1$).}
\label{fig:noise}
\end{figure}

\noindent For completeness, we next examine the robustness of the reconstruction algorithm with respect to random modeling errors. To this end, we assume the ``true'' material parameters to vary  (from one finite element to another) according to
\begin{equation}
\mu_\eta = \mu (1 + \eta \tau)\;, \quad \lambda_\eta = \lambda (1 + \eta \tau) \quad \text{and} \quad \rho_\eta = \rho (1 + \eta \tau) \;,
\label{eq:noisydata}
\end{equation}
where $\tau:\Omega \mapsto (0,1)$ is a random variable, $\eta$ specifies the amplitude of fluctuations and the domain is subdivided into $10\times 10$ subregions. To have a meaningful representation of material heterogeneities, each subregion is discretized by $4^4$ triangular elements where the corrupted material parameters are evaluated according to \eqref{eq:noisydata}. In this way, the average heterogeneity size~$d_h$ can be computed as $d_h/\lambda_s = (5/10)/4^2\simeq 0.03$, i.e. 3\% of the shear wavelength. For consistency, such material distribution is then projected onto a finer mesh with $4^7$ triangular elements per subregion, leading to a finite element discretization that is commensurate with those in Sections~\ref{1}--\ref{3}. As before, the reconstruction algorithm assumes a homogeneous background model with Lam\'{e} parameters $\lambda=\mu$ and mass density~$\rho$. For completeness, the perturbation function $(1 + \eta \tau)$ is plotted in Fig.~\ref{fig:noise} with $\eta=1$. 

\begin{remark}
With reference to~\eqref{eq:noisydata}, we note that the assumed perturbation does not affect the phase velocity in the elastic solid, since for instance we have $c_s=\sqrt{\mu/\rho} = \sqrt{\mu_\eta/\rho_\eta}=c_{s,\eta}$ in terms of shear waves. Such fluctuation, however, does affect the seismic impedance inside~$\Omega$; for example it is clear that $\rho\,c_s\neq \rho_\eta\, c_{s,\eta}$, which inherently affects the elastic wave reflection and transmission between neighboring finite elements.
\end{remark}

\begin{figure}[htbp]
\begin{center}\vspace*{3mm}
\subfigure[Real part]{\includegraphics[scale = 0.8]{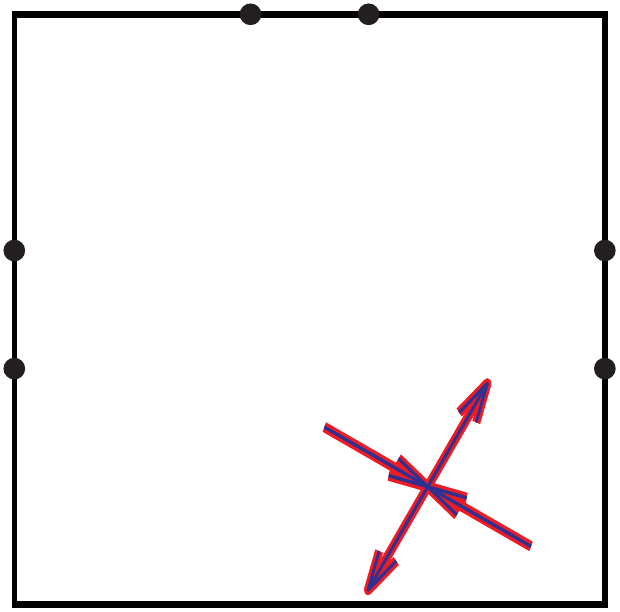}} \qquad
\subfigure[Imaginary part]{\includegraphics[scale = 0.8]{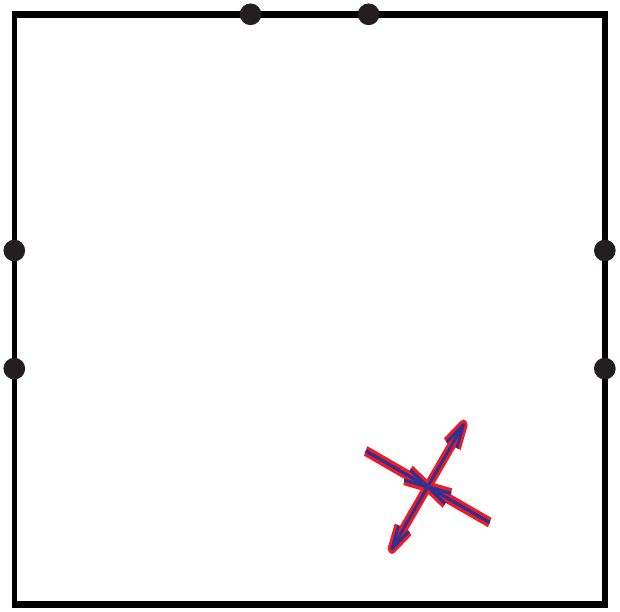}}
\end{center}\vspace*{-3mm}
\caption{Reconstruction of a single (mode II) micro-seismic source: background perturbation level $\eta = 0.0\%$ ($\mathcal{N}=0\%$ and $\mathcal{E}=0\%$).}
\label{fig:exe4a}
\end{figure}

\begin{figure}[htbp]
\begin{center}\vspace*{3mm}
\subfigure[Real part]{\includegraphics[scale = 0.8]{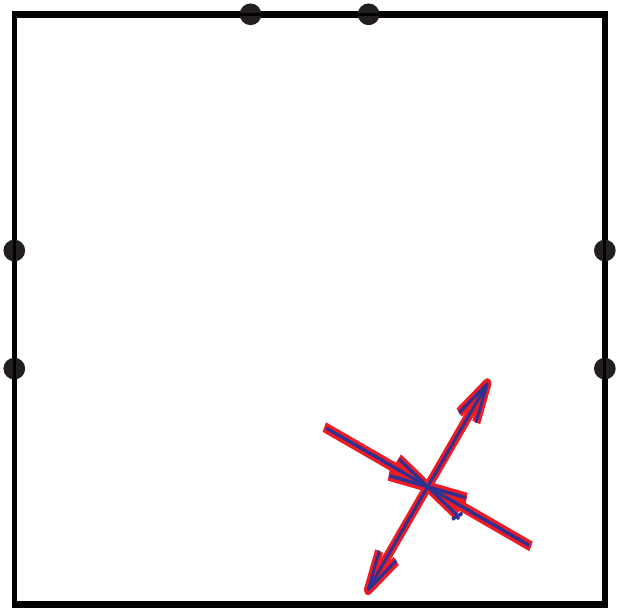}} \qquad
\subfigure[Imaginary part]{\includegraphics[scale = 0.8]{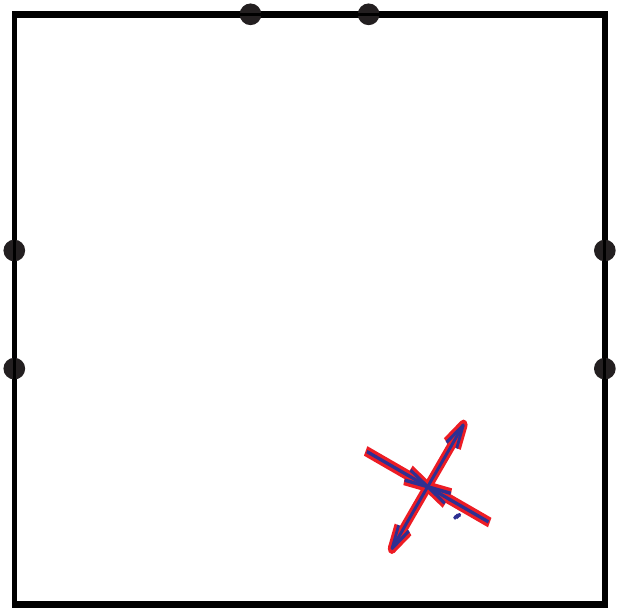}}
\end{center}\vspace*{-3mm}
\caption{Reconstruction of a single (mode II) micro-seismic source: background perturbation level $\eta = 0.5\%$ ($\mathcal{N}=10\%$ and $\mathcal{E}=2\%$).}
\label{fig:exe4b}
\end{figure}

\begin{figure}[htbp]
\begin{center}\vspace*{3mm}
\subfigure[Real part]{\includegraphics[scale = 0.8]{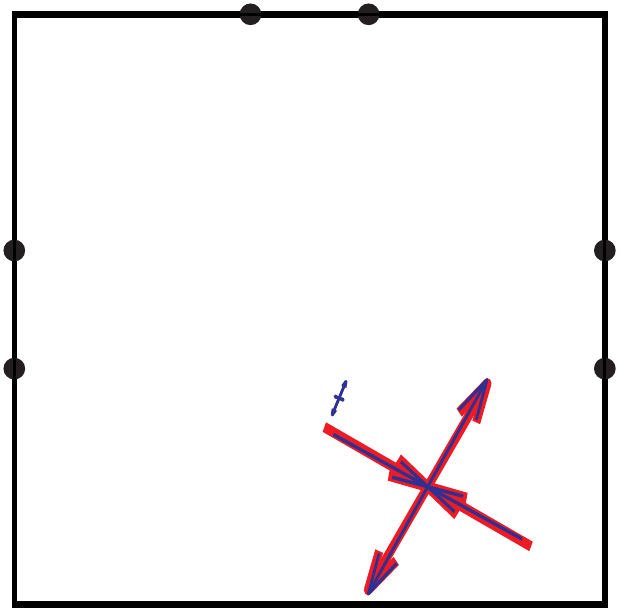}} \qquad
\subfigure[Imaginary part]{\includegraphics[scale = 0.8]{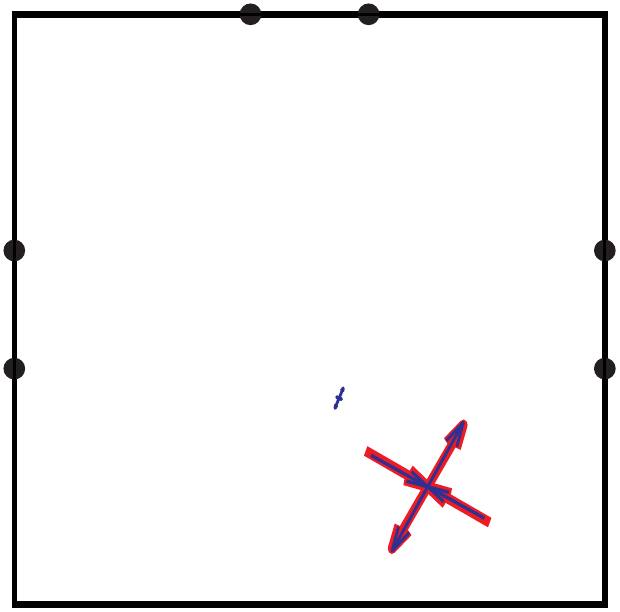}}
\end{center}\vspace*{-3mm}
\caption{Reconstruction of a single (mode II) micro-seismic source: background perturbation level $\eta = 1.0\%$ ($\mathcal{N}=21\%$ and $\mathcal{E}=11\%$).}
\label{fig:exe4c}
\end{figure}

\begin{figure}[htbp]
\begin{center}\vspace*{3mm}
\subfigure[Real part]{\includegraphics[scale = 0.8]{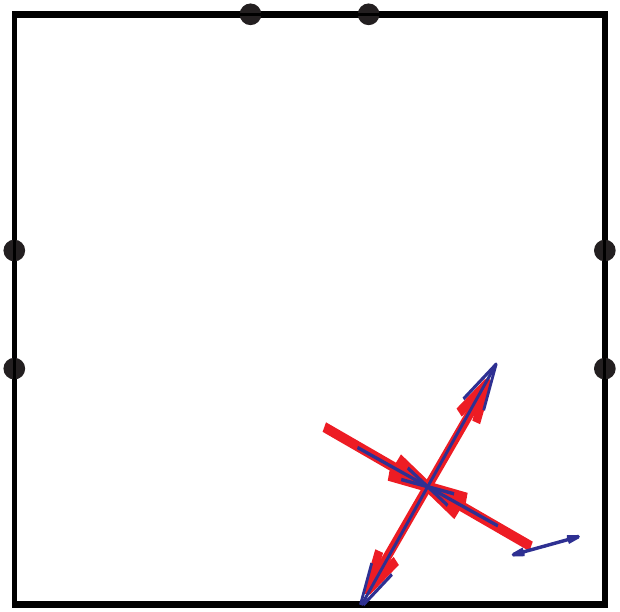}} \qquad
\subfigure[Imaginary part]{\includegraphics[scale = 0.8]{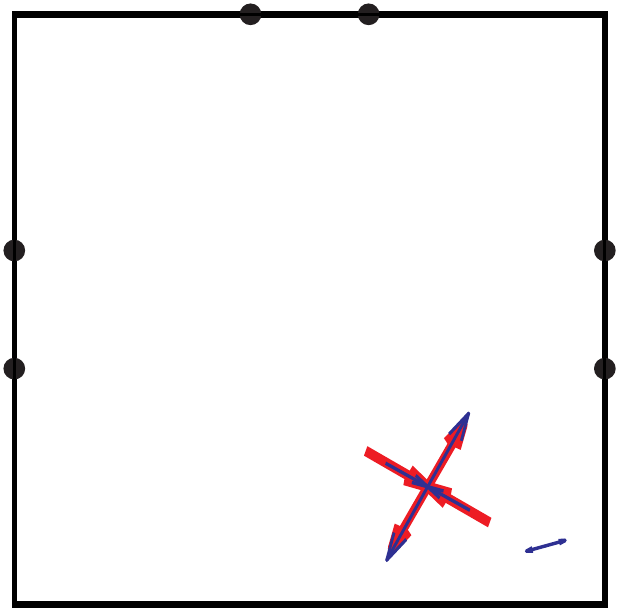}}
\end{center}\vspace*{-3mm}
\caption{Reconstruction of a single (mode II) micro-seismic source: background perturbation level $\eta = 2.0\%$ ($\mathcal{N}=51\%$ and $\mathcal{E}=33\%$).}
\label{fig:exe4d}
\end{figure}

In the first example, we aim to reconstruct a single mode II event with {\color{black}$\bar{\gamma}_{(1)} = 0.05 + 0.03 \mathfrak{i}$} and $\theta_{(1)} = 15^{\circ}$ using the six sensors shown in Fig.~\ref{fig:exe3}. The coordinates of the microcrack and those of the sensors are given in Table \ref{tab:ex1:c}. We assume that the number of faults is not known, and we set $N=2>N^*=1$. The results of source reconstruction for $\eta=0.0\%, 0.5\%, 1.0\%$ and $2.0\%$ are shown respectively in Figs.~\ref{fig:exe4a}--\ref{fig:exe4d}. For $\eta = 0.0\%$, the reconstruction is nearly exact. For $\eta = 0.5\%$, the reconstruction is still good, but there is a minuscule artifact in the form of a ``phantom'' second event as permitted by the premise $N=2$. This type of solution degradation continues to grow for $\eta = 1.5\%$ and $\eta = 2.0\%$ as can be seen from the respective displays. 

For completeness of discussion, we next introduce the effective ``noise level'' in the data due to~\eqref{eq:noisydata} as 
\begin{equation}
\mathcal{N} := \frac{\|\boldsymbol{u}_0 -\boldsymbol{u}_\eta\|_{L^2(\Omega)}}{\|\boldsymbol{u}_0\|_{L^2(\Omega)}},
\end{equation}
where $\boldsymbol{u}_0={\boldsymbol{u}_\eta}_{|_{\eta=0}}$ and $\boldsymbol{u}_\eta$ is the acoustic emission field due to exact source distribution~\eqref{truef} computed assuming~\eqref{eq:noisydata} for the background solid. Similarly, we introduce the resulting error in the reconstruction of the moment tensor as 
\begin{equation}
\mathcal{E} := \dfrac{\|  \boldsymbol{M}^\ast_{(1)} - \boldsymbol{M}_{(1)} \|}{\|\boldsymbol{M}^\ast_{(1)}\|},
\end{equation}
where $\|\boldsymbol{\cdot}\|$ denotes the Frobenius norm. With such definitions at hand, Table \ref{tab:exe4} lists~$\mathcal{N}$ and~$\mathcal{E}$ for $\eta = 0.5\%, 1.0\%$ and $2.0\%$. As can be seen from the tabulated values, the moment tensor reconstruction is fairly resilient to ``noise'' present in the data.

\begin{table}[htbp]
\centering
\caption{Reconstruction of a single (mode II) micro-seismic source: ``Noise level" in the data and relative error in the reconstruction of the moment tensor versus background perturbation level. \label{tab:exe4}} \vspace*{4pt} 
{\color{black}
\begin{tabu}{ccc}
\tabucline[1 pt]{-}
$\eta$     &  $\mathcal{N}$     &   $\mathcal{E}$    \\ \tabucline[1 pt]{-}
$0.5\%$             & $10\%$     &    2\% \\
$1.0\%$             & $21\%$     &    11\% \\
$2.0\%$             & $51\%$     &   33\%  \\
\tabucline[1 pt]{-}
\end{tabu}}
\end{table}

\begin{figure}[htbp]
\begin{center}\vspace*{3mm}
\subfigure[Real part]{\includegraphics[scale = 0.8]{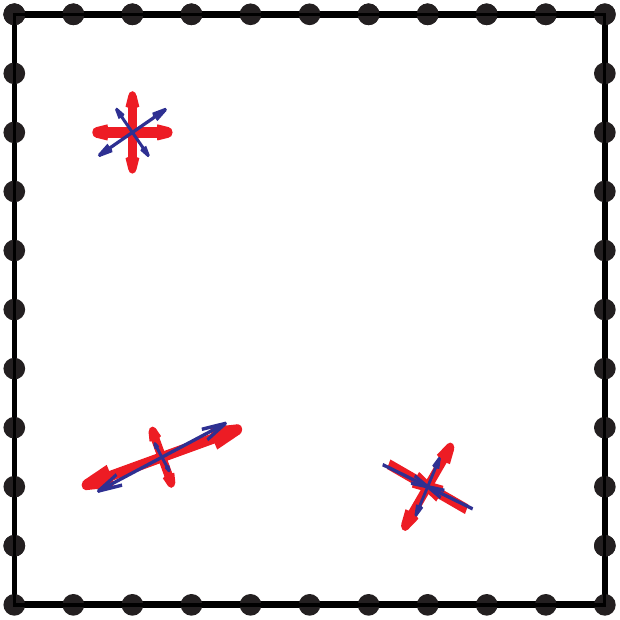}} \qquad
\subfigure[Imaginary part]{\includegraphics[scale = 0.8]{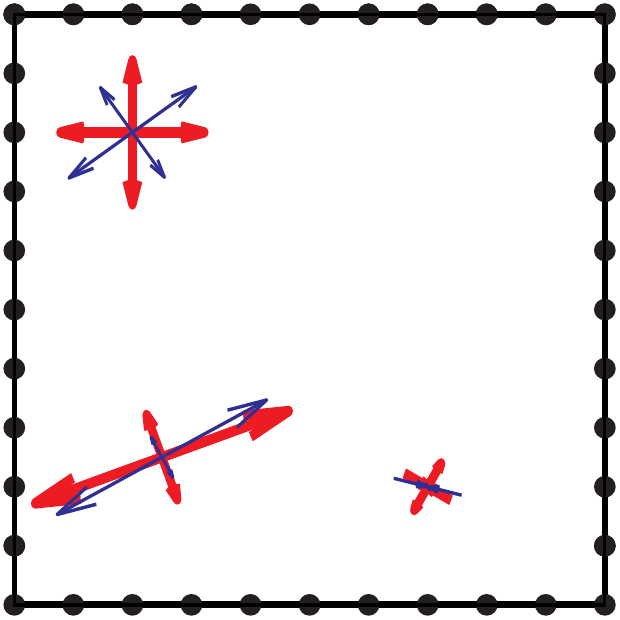}}
\end{center}\vspace*{-3mm}
\caption{Reconstruction of a triplet of micro-seismic sources: background perturbation level $\eta = 1.0\%$ ($\mathcal{N} = 21\%$).}
\label{fig:exe5a}
\end{figure}

\begin{figure}[htbp]
\begin{center}\vspace*{3mm}
\subfigure[Real part]{\includegraphics[scale = 0.8]{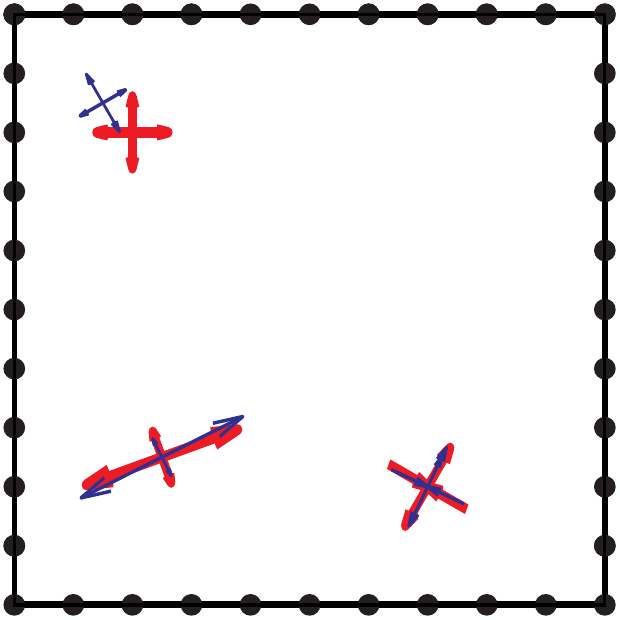}} \qquad
\subfigure[Imaginary part]{\includegraphics[scale = 0.8]{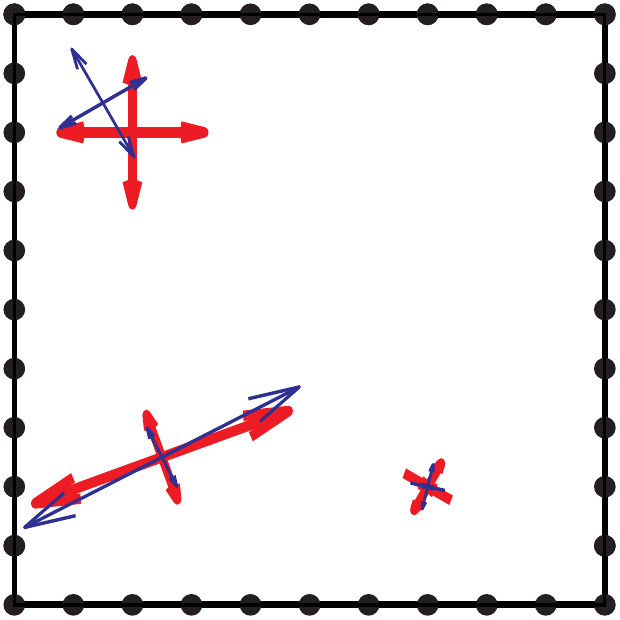}}
\end{center}\vspace*{-3mm}
\caption{Reconstruction of a triplet of micro-seismic sources: background perturbation level  $\eta = 1.5\%$ ($\mathcal{N} = 36\%$).}
\label{fig:exe5b}
\end{figure}

In the last example, we aim to reconstruct three co-existing events in the perturbed medium~\eqref{eq:noisydata} by setting $N = N^* = 3$. The target is the same as in Section~\ref{3}, see Table~\ref{tab:ex1:c} for event locations. In this case, however, we use 40 sensors uniformly distributed on~$\partial\Omega$ in order to combat the modelling errors. For $\eta = 0.0\%$, the reconstruction is nearly exact and practically the same as in Fig.~\ref{fig:exe3}. The source reconstructions obtained for $\eta = 1.0\%$ and $\eta = 1.5\%$ are shown respectively in Fig.~\ref{fig:exe5a} and  Fig.~\ref{fig:exe5b}. For $\eta = 1.0\%$, the result is still reasonable in the sense that (i) the event locations are accurately resolved and (ii) the character of each event is preserved (cavitation vs. mode I crack vs. mode II crack), despite apparent degradation in the moment tensor reconstruction. However, for $\eta = 1.5\%$ the reconstruction error is significant in that the algorithm is unable to resolve the cavitation event near the upper left corner of the domain.

% --------------------------------------------------------------------------
\section{Conclusions} \label{sec:conclu}
% --------------------------------------------------------------------------

\noindent In this study, we propose an algorithm for the frequency-domain reconstruction of \mbox{(micro-)} seismic events using full-waveform analysis of the acoustic emission data. The inversion approach integrates a combinatorial grid search for source locations with the sensitivity analysis in terms of moment tensor components to arrive at an effective algorithm that simultaneously returns both micro-seismic source coordinates and respective tensorial ``strengths''. We investigate the performance of the algorithm, assuming pointwise waveform observations, via numerical examples that include both isolated and multiple point sources. Under ideal testing conditions, the results suggest that two point receivers per acoustic emission source may provide sufficient information for accurate inversion. To enable the reconstruction of arbitrarily located (``off-grid") sources, we also introduce an iterative scheme that recursively refines the search grid around ``coarsely" reconstructed source locations. The results show that the course reconstructions are inherently confined to the neigborhood of ``true'' source locations, thus lending credence to the proposed recursive scheme. For generality, we also investigate the micro-seismic source reconstruction under the adverse condition of randomly perturbed background medium, whose local fluctuations are unavailable as prior information. The results show a significant resilience of the reconstruction algorithm to this type of modeling errors.

% --------------------------------------------------------------------------
\section*{Acknowledgements}
% --------------------------------------------------------------------------

\noindent This research was partly supported by CNPq (Brazilian Research Council), CAPES (Brazilian Higher Education Staff Training Agency) and FAPERJ (Research Foundation of the State of Rio de Janeiro). The support is gratefully acknowledged. The third author kindly acknowledges the support provided by the endowed Shimizu Professorship and the U.S. National Science Foundation (CMMI Grant \#1536110) during the course of this investigation.

% --------------------------------------------------------------------------
\section*{References}
% --------------------------------------------------------------------------

%--------------------------
\bibliographystyle{apalike}
\bibliography{BibLNCC,AE}

\begin{thebibliography}{}

\bibitem[Aki and Richards, 2002]{Aki2002}
Aki, K. and Richards, P. (2002).
\newblock {\em Quantitative Seismology}.
\newblock Sausalito, Calif, University Science Books.

\bibitem[Baig and Urbancic, 2010]{Baig2010}
Baig, A. and Urbancic, T. (2010).
\newblock Microseismic moment tensors: A path to understanding frac growth.
\newblock {\em The Leading Edge}, 29:320--324.

\bibitem[Bazargani and Snieder, 2015]{Baza2015}
Bazargani, F. and Snieder, R. (2015).
\newblock Optimal source imaging in elastic media.
\newblock {\em Geophys. J. Int.}, 204:1134--1147.

\bibitem[Canelas et~al., 2014]{CanelasJCP2014}
Canelas, A., Laurain, A., and Novotny, A.~A. (2014).
\newblock A new reconstruction method for the inverse potential problem.
\newblock {\em Journal of Computational Physics}, 268:417--431.

\bibitem[Cesca and Dahm, 2008]{Cesc2008}
Cesca, S. and Dahm, T. (2008).
\newblock A frequency domain inversion code to retrieve time-dependent
  parameters of very long period volcanic sources.
\newblock {\em Computers \& Geosciences}, 34:235--246.

\bibitem[Cesca and Grigoli, 2015]{Cesc2015}
Cesca, S. and Grigoli, F. (2015).
\newblock Chapter two - full waveform seismological advances for microseismic
  monitoring.
\newblock volume~56 of {\em Advances in Geophysics}, pages 169--228. Elsevier.

\bibitem[Gilbert, 1971]{Gilb1971}
Gilbert, F. (1971).
\newblock Excitation of the normal modes of the earth by earthquake sources.
\newblock {\em Geophys. J. Int.}, 22:223--226.

\bibitem[Grosse and Ohtsu, 2008]{Gros2008}
Grosse, C. and Ohtsu, M. (2008).
\newblock {\em Acoustic Emission Testing}.
\newblock Springer Science \& Business Media.

\bibitem[Jost and Herrmann, 1989]{Jost1989}
Jost, M. and Herrmann, R. (1989).
\newblock A student’s guide to and review of moment tensors.
\newblock {\em Seism. Res. Lett.}, 60:37--57.

\bibitem[Kawakatsu and Montagner, 2008]{Kawa2008}
Kawakatsu, H. and Montagner, J.-P. (2008).
\newblock Time-reversal seismic-source imaging and moment-tensor inversion.
\newblock {\em Geophys. J. Int.}, 175:686--688.

\bibitem[Koerner et~al., 1981]{Koer1981}
Koerner, R., McCabe, W., and Lord, A. (1981).
\newblock Overview of acoustic emission monitoring of rock structures.
\newblock {\em Rock Mechanics}, 14:27--35.

\bibitem[Novotny and Soko{\l}owski, 2013]{NovotnyBook2013}
Novotny, A.~A. and Soko{\l}owski, J. (2013).
\newblock {\em Topological derivatives in shape optimization}.
\newblock Interaction of Mechanics and Mathematics. Springer-Verlag, Berlin,
  Heidelberg.

\bibitem[Novotny et~al., 2019a]{NovotnyBook2019}
Novotny, A.~A., Soko{\l}owski, J., and {\.Z}ochowski, A. (2019a).
\newblock {\em Applications of the topological derivative method}.
\newblock Studies in Systems, Decision and Control. Springer Nature
  Switzerland.

\bibitem[Novotny et~al., 2019b]{NovotnyJOTA2019c}
Novotny, A.~A., Soko{\l}owski, J., and {\.{Z}}ochowski, A. (2019b).
\newblock Topological derivatives of shape functionals. {Part III}: Second
  order method and applications.
\newblock {\em Journal of Optimization Theory and Applications}, 181(1):1--22.

\bibitem[Rice, 1980]{Rice1980}
Rice, J. (1980).
\newblock Elastic wave emission from damage processes.
\newblock {\em J. Nondestr. Eval.}, 1:215--224.

\bibitem[Scruby et~al., 1985]{Scru1985}
Scruby, C., Baldwin, G., and Stacey, K. (1985).
\newblock Characterisation of fatigue crack extension by quantitative acoustic
  emission.
\newblock {\em Int. J. Fracture}, 28:201--222.

\bibitem[Shearer, 2009]{Shea2009}
Shearer, P. (2009).
\newblock {\em Introduction to Seismology}.
\newblock Cambridge University Press.

\bibitem[Sj{\"o}green and Petersson, 2014]{Sjor2014}
Sj{\"o}green, B. and Petersson, N. (2014).
\newblock Source estimation by full wave form inversion.
\newblock {\em J. Sci. Comput.}, 59(1):247--276.

\bibitem[Song and Toks{\"o}z, 2011]{Song2011}
Song, F. and Toks{\"o}z, M. (2011).
\newblock Full-waveform based complete moment tensor inversion and source
  parameter estimation from downhole microseismic data for hydrofracture
  monitoring.
\newblock {\em Geophysics}, 76:WC103--WC116.

\end{thebibliography}
%--------------------------

\end{document}